\documentclass[11pt]{article}

\usepackage{amssymb}
\usepackage{subfig}
\usepackage{color}
\usepackage{epsfig,amssymb,amsfonts,amsmath,graphicx,dsfont,cite,xfrac}
\usepackage{authblk}
\definecolor{mygray}{gray}{0.5}
\usepackage{cite}
\usepackage[colorlinks=true,linkcolor=blue,citecolor=red]{hyperref}

\usepackage{tikz,pgfplots}
\usetikzlibrary{decorations.pathreplacing}
\usetikzlibrary{shapes.multipart}
\usetikzlibrary{positioning}
\usetikzlibrary{matrix}
\usetikzlibrary{shapes,calc,arrows}

\newcommand{\be}{\begin{equation}}
\newcommand{\ee}{\end{equation}}
\newcommand{\bea}{\begin{eqnarray}}
\newcommand{\eea}{\end{eqnarray}}

\title{Form-preserving Darboux transformations for $4\times 4$ Dirac equations}

\author[${}$]{M. Castillo-Celeita, V. Jakubsk\'y, K. Zelaya  }

\affil[${}$]{\footnotesize Nuclear Physics Institute, Czech Academy of Science, Prague/\v{R}e\v{z}, Czech Republic}

\date{}

\begin{document}

\maketitle

\begin{abstract}
Darboux transformation is a powerful tool for the construction of new solvable models in quantum mechanics. In this article, we discuss its use in the context of physical systems described by $4\times4$ Dirac Hamiltonians. The general framework provides limited control over the resulting energy operator, so that it can fail to have the required physical interpretation. We show that this problem can be circumvented with the reducible Darboux transformation that can preserve the required form of physical interactions by construction. To demonstrate it explicitly, we focus on distortion scattering and spin-orbit interaction of Dirac fermions in graphene. We use the reducible Darboux transformation to construct exactly solvable models of these systems where backscattering is absent, i.e. the models are reflectionless.

\end{abstract}

\section{Introduction}
The Dirac equation rules relativistic dynamics of spin-$\frac{1}{2}$ quantum particles, which predicts the existence of antiparticles~\cite{Dir28}, hyperfine structure in the hydrogen atom and other muonic atoms~\cite{Bjo64,Gre00}, among other phenomena. Notably, the one- and two-dimensional Dirac equations emerge naturally in other physical scenarios 
such as optics~\cite{Bia94,Bar14,Hor18,Longhi,CorreaJakubsky}, geometrical and topological phases~\cite{Koi94,Lin14,Lon21}, graphene layers~\cite{Cas09,Cas21,Can06,Can13}, and Bose-Einstein condensates~\cite{Had11}. In these scenarios, predictions obtained from the Dirac equation can be tested in non-relativistic setups, which are usually more accessible to implement, such as in microwave cavities~\cite{Sad10,Fra13}.

From the mathematical-physics perspective, solvable models play an essential role by providing further insight into physical phenomena. Exact solutions grant us global information of the system and are by construction free of error. This is in contrast to numerical approaches, where numerical solutions bring only local information and have intrinsic error. On the other hand, solvable models provide a test field for computational methods, and they can serve in perturbation analysis as the initial unperturbed system. In this regard, supersymmetric quantum mechanics offers us effective tools for construction of exactly solvable models of both non-relativistic and relativistic systems \cite{CooperKhare, Junker, Junker2}. Darboux transformation forms the backbone of supersymmetric quantum mechanics. In the context of the Dirac equation, there are two possible implementations followed in the literature. 

In the first approach, Darboux transformation is not applied directly on the Dirac Hamiltonian, but rather on the Schr\"odinger-like operator that corresponds to the square of Dirac Hamiltonian, see e.g. \cite{Kuru,Contreras1,Contreras2,Midya,ozlem,Celeita,fernandez21}. In the second case, Darboux transformation is used directly to modify Dirac energy operators.  For $2\times2$ Dirac operator\footnote{By $2\times2$ operator we denote an operator acting on $\mathbb{C}^{2\times2}\otimes \mathcal{H}$ where $\mathcal{H}$ is a Hilbert space.}, it was discussed in~\cite{samsonovstac,samsonovtimedep}. For general one-dimensional Dirac operator given in terms of $n\times n$ matrices, it was discussed in~\cite{schulzehalberg}. This approach was utilized e.g. for construction of the new exactly solvable models based on Heun equation \cite{Ishkhanyan}, describing mechanical deformations of nanotubes \cite{JakubskyNietoPlyushchay,JakubskyPlyushchay,Cor14} or in the description of optical systems \cite{CorreaJakubsky,Cha21}. 

Darboux transformation can serve well in constructing a new Dirac Hamiltonian that describes an exactly solvable system. Yet, it can be challenging to obtain it in the required and physically relevant form. In the current article, we shall discuss some specific Darboux transformations for $4\times 4$ Dirac operators such that, after introducing some appropriate unitary transformations, we still keep control over the structure of the new Hamiltonian. This allows us to discuss some examples of interest in physics of graphene. 

The work is organized as follows. In section~\ref{sec:Dirac}, we review the Darboux transformation for a generic $m\times m$ Dirac operator, where we show that the general construction makes it challenging to keep control over the resulting model. In section \ref{section3}, we discuss a class of $4\times 4$ Dirac operators that are reducible. We show there that these operators allow us to use reduced or partial Darboux transformation in order to obtain energy operators for specific physical systems. In section \ref{sec:2x2-reflectionless}, we discuss in detail reflectionless systems described by $2\times 2$ Dirac Hamiltonian. These results are used directly in the section \ref{section5}, where solvable, reflectionless systems with distortion scattering or spin-orbit interaction are constructed. We discuss possible caveats that emerge in case of non-reducible Darboux transformation in section \ref{sec:non-red}. The last section is left for summary and discussion.

\section{Dirac equation and Darboux transformations}
\label{sec:Dirac}
To begin with, let us introduce some notation and definitions. We will deal with quantum systems described by the \textit{stationary Dirac equation}
\begin{equation}
H\mathbf{\Psi}=(\gamma \partial_{x}+V(x))\mathbf{\Psi}=E\mathbf{\Psi} \, , 
\label{dir2}
\end{equation}
where $\gamma,V(x)\in\mathbb{C}^{m\times m}$ and $m\in\mathbb{Z}^{+}$.
We require $H$ to be hermitian in $\mathbb{C}^{m}\otimes L^{2}(\mathbb{R};dx)$, which we henceforth denote by $\mathbb{C}^{m}\otimes{L}^{2}$ for short. It implies that there should hold $\gamma^{\dagger}=-\gamma$ and $V(x)^{\dagger}=V(x)$.  The scalar product on $\mathbb{C}^{m}\otimes{L}^{2}$ is defined in standard manner, so that for two $m$-component solutions $\mathbf{\Psi}=(\psi_{1},\ldots,\psi_{m})^{T}$ and $\mathbf{\Phi}=({\phi}_{1},\ldots,{\phi}_{m})^{T}$ of (\ref{dir2}), the scalar product reads as
\begin{equation}
({\mathbf{\Psi}},\mathbf{\Phi})\equiv 
\int_{\mathcal{\mathbb{R}}}dx\mathbf{\Psi}^{\dagger}\mathbf{\Phi}=\int_{\mathbb{R}}dx \, \sum_{k=1}^{m}{\psi}^{*}_{k}(x)\phi_{k}(x) \, .
\label{inner}
\end{equation}
The corresponding probability density is 
\begin{equation}
\mathcal{P}_{\mathbf{\Psi}}:=(\psi_{1}^{*},\ldots,\psi_{m}^{*})\cdot(\psi_{1},\ldots,\psi_{m})=\sum_{k=1}^{m}\vert\psi_{k}\vert^{2} \, ,
\end{equation}
and the corresponding norm is defined through the scalar product as $\Vert \mathbf{\Psi} \Vert^{2}=(\boldsymbol{\Psi},\boldsymbol{\Psi})$. We say that $\boldsymbol{\Psi}$ is a bound state if it fulfills~\eqref{dir2} and has finite norm, $\Vert\boldsymbol{\Psi}\Vert<\infty$. We also use the term ``square integrable'' interchangeably.

\subsection{Darboux transformations}

Now, let us briefly review the Darboux transformation for Dirac equation (\ref{dir2}). We refer to \cite{samsonovstac} and \cite{schulzehalberg} for more details. Darboux transformation is typically represented by a differential operator that mediates intertwining relations between two Dirac Hamiltonians $H$ and $\widetilde{H}$,
\begin{equation}
\mathcal{L} H=\widetilde{H}\mathcal{L} \, .
\label{intert1}
\end{equation}
When an eigenstate $\mathbf{\Psi}$ of $H$ is known, then (\ref{intert1}) implies that we can find an eigenstate $\mathcal{L}\mathbf{\Psi}$ of $\widetilde{H}$ corresponding to the same eigenvalue, except the cases where $\mathcal{L}\mathbf{\Psi}=0$. Conjugating (\ref{intert1}), one can check that $\mathcal{L}^{\dagger}$ does the converse provided that both $H$ and $\widetilde{H}$ are hermitian. Although $\mathcal{L}^{\dagger}$ reverts the action of $\mathcal{L}$, it is not the inverse of $\mathcal{L}$. Let us show how $\mathcal{L}$ and $\widetilde{H}$ can be found such that (\ref{intert1}) is satisfied for a given Hamiltonian $H$. 

Let us suppose that we have two stationary Dirac equations of the form
\begin{eqnarray}
&& H\mathbf{\Phi}\equiv\left(\gamma\partial_{x}+V(x)\right)\mathbf{\Phi}= E\mathbf{\Phi} \, , \\ 
&& \widetilde{H}\widetilde{\mathbf{\Phi}}\equiv\left(\gamma\partial_{x}+\widetilde{V}(x) \right)\widetilde{\mathbf{\Phi}}=\widetilde{E}\widetilde{\mathbf{\Phi}} \, ,
\label{mathcalH}
\end{eqnarray}
with $\gamma^{\dagger}=-\gamma$ an arbitrary invertible $m\times m$ constant matrix and  $V^{\dagger}=V$ and $\widetilde{V}^{\dagger}=\widetilde{V}$ the corresponding matrix potentials. We suppose that the Hamiltonian $H$ is given explicitly whereas $\mathcal{L}$ and $\widetilde{V}$ are to be fixed such that (\ref{intert1}) is satisfied. 
We make an anstatz for $\mathcal{L}$ in terms of the first-order differential operator of the form\footnote{Strictly speaking, the intertwining operator should write $\mathcal{L}=\mathbb{I}_{m\times m}\partial_{x}-U_{x}U^{-1}$ with $\mathbb{I}_{m\times m}$ the identity matrix of dimension $m$. Nevertheless, for simplicity, we dropout the identity matrix form the notation.}
\begin{equation}
\mathcal{L}=\partial_x-U_{x}U^{-1}=U\partial_xU^{-1} \, ,
\label{L}
\end{equation}
where the $m\times m$ matrix $U$ is to be specified, and we have used the subindex notation $f_{x}=\partial_{x}f$. This formula is inspired by supersymmetry in nonrelativistic quantum mechanics and was introduced in~\cite{samsonovstac}. Eq.~\eqref{L} facilitates better insight into the structure of the Darboux transformation. Indeed, substituting (\ref{L}) into~(\ref{intert1}) and comparing the coefficients of the terms with the same order of derivatives, we get the following two equations
\begin{eqnarray}
\widetilde{V}=V+[\gamma,U_{x}U^{-1}], \quad V_{x}-U_{x}U^{-1}V=-\gamma (U_{x}U^{-1})_{x}-\widetilde{V}U_{x}U^{-1} \, .
\label{intert2}
\end{eqnarray}
The first equation identifies the new potential $\widetilde{V}$ in terms of $V$ and $U$ whereas the second one determines the matrix $U$. If we substitute the first equation in~\eqref{intert2} into the second one,  multiply the resulting equation by $U^{-1}$ to the left, and integrate with respect to $x$, we obtain
\begin{equation}
\left(\gamma\partial_{x}+V\right)U=U\Lambda \, , \quad \Lambda=\operatorname{diag}(\epsilon_{1},\ldots,\epsilon_{m}) \, .
\label{eigenU}
\end{equation}
Here the constant matrix $\Lambda$ emerges as an integration constant. For the sake of simplicity, we assume it to be diagonal. From~\eqref{eigenU}, it is clear that $U$ is composed by the eigensolutions $\mathbf{\Phi}_{k}$ of $H$, $H\mathbf{\Phi}_{k}=\epsilon_k\mathbf{\Phi}_{k}$, for $k=1,\ldots,m$. Explicitly we have 
\begin{equation}
U=(\mathbf{\Phi}_{k},\ldots,\mathbf{\Phi}_{m}) \, , \quad \mathbf{\Phi}_{k}=(\phi_{1;k},\ldots,\phi_{m;k})^{T} \, , \quad k=1,\ldots,m\, .
\label{U}
\end{equation}
The eigensolutions $\mathbf{\Phi}_{k}$ (sometimes called ``see'' solutions in the literature) are not necessarily square integrable. Notice that there holds $\mathcal{L}\mathbf{\Phi}_{k}=0$ by definition.  

The definition (\ref{U}) completes the construction of $\widetilde{H}$ and $\mathcal{L}$. Indeed, when we select the eigenstates $\mathbf{\Phi_k}$ in (\ref{U}), we can compose the matrix $U$. Then we can find the new Hamiltonian $\widetilde{H}$ and the intertwining operator $\mathcal{L}$ in terms of this matrix, see (\ref{L}) and (\ref{intert2}).  We have to stress that the construction is formal.  It is not guaranteed that the action of $\mathcal{L}$ preserves the required boundary conditions, i.e. it is not guaranteed that both $\mathbf{\Phi}$ and $\mathcal{L}\mathbf{\Phi}$ correspond to physical states. Additionally, it is not guaranteed that the new potential term $\widetilde{V}$ is hermitian, and it may even have additional singularities when $\det (U)=0$ for some $x$. The properties of $\widetilde{H}$ are determined by choice of $U$, which should be fixed such that the new Hamiltonian has the required form. This task can be very demanding in general and it will be one of the main topics of this article.

When both $H$ and $\widetilde{H}$ are hermitian, they also satisfy $H\mathcal{L}^\dagger=\mathcal{L}^\dagger\widetilde{H}$. We can see from (\ref{L}) that $\mathcal{L}^\dagger$ annihilates the matrix $\widetilde{U}=(U^{-1})^\dagger$,
\begin{equation}
\mathcal{L}^\dagger\equiv -\widetilde{U}\partial_x\widetilde{U}^{-1}=-(U^{-1})^\dagger\partial_x U^\dagger,
\end{equation}
\begin{equation}
\widetilde{U}\equiv(\widetilde{\mathbf{\Phi}}_{\epsilon_1},\ldots,\widetilde{\mathbf{\Phi}}_{\epsilon_m})\ ,\quad \widetilde{H}\widetilde{\mathbf{\Phi}}_{\epsilon_k}=\epsilon_k \widetilde{\mathbf{\Phi}}_{\epsilon_k}, \quad k\in\{1,\dots, m\}.
\label{missing}
\end{equation}
The states $\widetilde{\mathbf{\Phi}}_{\epsilon_k}$ are frequently called ``missing'' eigenstates in the literature. It is not guaranteed that these states are physically acceptable. It has to be checked that they comply with the required boundary conditions. In this article, we will deal with the situation where the energy levels $\epsilon_{k}$ are absent in the spectrum of $H$, i.e. $\mathbf{\Phi}_{\epsilon_k}$ in (\ref{U}) are not square integrable, nevertheless, $\widetilde{\mathbf{\Phi}}_{\epsilon_k}$ have finite norm and $\epsilon_{k}$ form discrete energies of $\widetilde{H}$. 

\subsection{Challenges of $4\times 4$ Darboux transformation and physics of graphene}

The relations (\ref{intert1})-(\ref{U}) provide a powerful tool to construct solvable equations of the form (\ref{dir2}) with $\widetilde{V}$ as the new potential term. Yet, the physical interpretation of such equation can be complicated.  We are primarily interested in physics of graphene where (\ref{dir2}) appears in low-energy description of quasi-particles. The dimension of the matrix coefficients is related to the number of degrees of freedom that are considered in the system. In general, there is pseudo-spin degree of freedom induced by presence of two atoms in the elementary cell of graphene crystal. The valley degree of freedom is related to existence of two inequivalent Dirac points in the first Brillouin zone of graphene. Finally, there is also a spin degree of freedom of the electrons. Therefore, the Hamiltonian reflecting all the properties would acquire the form of a $\mathbb{C}^{8\times 8}$ matrix, i.e. $m=8$ in (\ref{dir2}). Nevertheless, only those degrees of freedom relevant for the considered interaction appear in the effective description. 

Throughout this manuscript, we will be primarily interested the situations where the effective Hamiltonian is given in terms of $4\times 4$ matrices, i.e. $m=4$ in (\ref{dir2}). 
Such energy operator can describe distortion scattering or spin-orbit interaction of Dirac fermions in graphene. Let us consider the following two types of energy operators, 
%
%
%
\begin{eqnarray}
H_{dis}&=&-i\sigma_3\otimes\sigma_1\partial_x+V_{dis}=\left(\begin{array}{cccc}
V_A&-i\partial_x+V&W_A&W^{+}\\
-i\partial_x+{V}^*& V_B&W^{-}&W_B\\
(W_A)^{*} & (W^{-})^{*} & V_A & i\partial_x+V'\\
(W^{+})^{*} & (W_B)^{*} & i\partial_x+(V')^{*}&V_B
\end{array}\right),\nonumber\\
H_{soc}&=&-i\sigma_0\otimes\sigma_1\partial_x+V_{soc}=\left(\begin{array}{cccc}V-
\Delta&-i\partial_x&0&0\\
-i\partial_x&V+\Delta&-2i\lambda&0\\
0&2i\lambda&V+\Delta&-i\partial_x\\
0&0&-i\partial_x&V-\Delta
\end{array}\right).
\label{Hsoc}
\end{eqnarray}
The Hamiltonian $H_{dis}$ of disorder scattering acts on the bispinors whose components are ordered as $(\psi_A^K,\psi_B^K,\psi_A^{K'},\psi_B^{K'})$, where $\psi_{A(B)}^{K(K')}$ is the wave function localized at the lattice $A(B)$ and describing states from the vicinity of the Dirac point $K(K')$, see e.g. ~\cite{Can06,Can13}. The interactions associated with $W_A$, $W_B$ and $W^\pm$ cause intervalley scattering, which may be caused by impurities or atomic defects of the crystal lattice of mechanical contact with substrate \cite{Altland}, \cite{Kechedzhi}, \cite{Manes}.
The Hamiltonian $H_{soc}$ of spin-orbit interaction acts on the bispinors that have the following ordering $(\psi_A^\uparrow,\psi_B^\uparrow,\psi_A^{\downarrow},\psi_B^{\downarrow})$. Here $\psi_{A(B)}^{\uparrow(\downarrow)}$ is wave function localized at the lattice $A(B)$ and with spin up(down). The potential term $\Delta$ describes intrinsic spin-orbit interaction and $\lambda$ represents Rashba spin-orbit interaction \cite{Avsar}, \cite{Kane}, \cite{Huertas}. 

We would like to utilize Darboux transformation in the analysis of the systems described by (\ref{Hsoc}) by constructing solvable models with $\widetilde{H}\equiv H_{dis}$ or $\widetilde{H}\equiv H_{soc}$. As we mentioned in the previous subsection, this is a nontrivial task. Let us suppose that $H=\gamma\partial_x+V$ is hermitian and solutions of its stationary equation are known. Then, the new potential term $\widetilde{V}$ should be hermitian and free of additional singularities. The latter is achieved if 
\begin{equation}
[\gamma,U_xU^{-1}]^\dagger=[\gamma,U_xU^{-1}],\quad \det U\neq 0.
\label{hermitianregular}
\end{equation}
Additionally, $\widetilde{V}$ should have the same structure as either $V_{dis}$ or $V_{soc}$,  
\begin{equation}\label{identification}
\widetilde{V}=V(x)+[\gamma,U_xU^{-1}]\sim\begin{cases}V_{dis},\quad \gamma=i\sigma_3\otimes\sigma_1,\\
V_{soc},\quad \gamma=i\sigma_0\otimes\sigma_1.\end{cases}
\end{equation}
For instance, there should hold $\widetilde{V}_{12}=\widetilde{V}_{13}=\widetilde{V}_{14}=\widetilde{V}_{24}=\widetilde{V}_{34}=0$, $\widetilde{V}_{11}=\widetilde{V}_{44}$ and $\widetilde{V}_{22}=\widetilde{V}_{33}$ for $\widetilde{V}\sim V_{soc} $,
However, it is very nontrivial to fix $U$, i.e. to select the eigenstates $\mathbf{\Phi}_k$ in (\ref{U}),  such that (\ref{hermitianregular}) and (\ref{identification}) are satisfied. Each state $\mathbf{\Phi}_k$ is a linear combination $\mathbf{\Phi}_k=\sum_{j=1}^4c_{k;j}\mathbf{\Phi}_k^{(j)}$ of four fundamental solutions $\mathbf{\Phi}_k^{(j)}$ of the stationary equation $(H-\epsilon_k)\mathbf{\Phi}_k^{(j)}=0$. Thus, there are twenty parameters $c_{k;j}$ and $\epsilon_k$, $j,k=1,\dots,4$, in the matrix $U$ in general. In the relations  (\ref{hermitianregular}) and (\ref{identification}), there are rational functions composed by fourth-order polynomials $c_j$ in both the numerator and denominator. The eigenvalue $\epsilon_k$ is involved in the explicit form of the fundamental solutions $\mathbf{\Phi}_k^{(j)}$. 
These parameters have to be fine-tuned such that (\ref{hermitianregular}) and (\ref{identification}) are satisfied. We are not aware of any systematic way that would allow us to do it properly, or, in other words, to keep sufficient control over the form of $\widetilde{V}$. 

For the reasons mentioned above, direct use of Darboux transformation does not seem to be quite effective for construction of solvable systems where, besides Hermiticity, the Hamiltonian should take a specific form, e.g. in (\ref{Hsoc}). Nevertheless, let us show how the major problems can be circumvented.

\section{Reduction scheme and Darboux transformations\label{section3}}

There is a reducible class of Hamiltonians $H$ that can be brought into a block-diagonal form by implementing a unitary transformation generated by a unitary matrix $\mathcal{U}$ as
\begin{equation}
\mathcal{U}^{-1}H\mathcal{U}= \mathbb{S}_{1}\otimes \mathbf{h_{1}}+\mathbb{S}_{2}\otimes \mathbf{h_{2}} \, , \quad \mathcal{U}^\dagger=\mathcal{U}^{-1}.
\label{H-blocks}
\end{equation}
where the matrices $\mathbb{S}_1$ and $\mathbb{S}_2$ are defined as
\begin{equation}
	\mathbb{S}_{1}:=
	\begin{pmatrix}
		1 & 0 \\
		0 & 0
	\end{pmatrix}
	\, , \quad 
	\mathbb{S}_{2}:=
	\begin{pmatrix}
		0 & 0 \\
		0 & 1
	\end{pmatrix}
	\, , \quad \mathbb{S}_{j}\mathbb{S}_{k}=\delta_{j,k}\mathbb{S}_{j} \, .
	\label{s1s2}
\end{equation}
We suppose that the reduced operators $\mathbf{h_1}$ and $\mathbf{h_2}$ are hermitian
\begin{equation}
\mathbf{h_{j}}=-i\sigma_{1}\partial_{x}+\mathbf{v_j},
  \quad \mathbf{v_j}=\begin{pmatrix}
v_{j} & a_{j} \\
a_{j}^{*} & w_{j}
\end{pmatrix},
\, \quad \mathbf{v_j}^\dagger=\mathbf{v_j},\quad j=1,2,\label{h_a}
\end{equation}
so that $v_j(x),\,w_j(x)\in\mathbb{R} $ and $a_j(x)\in\mathbb{C}$, $j=1,2.$ 

Reducibility of $H$ simplifies the solutions of (\ref{dir2}) considerably. For instance, if one can find the spinors  $\boldsymbol{\xi}_E$ and $\boldsymbol{\chi}_E$ such that
\begin{equation}
(\mathbf{h_1}-E)\boldsymbol{\xi}_{E}=0,\quad (\mathbf{h_2}-E)\boldsymbol{\chi}_{E}=0,\quad \boldsymbol{\xi}_{E}=(\xi_{1;E},\xi_{2;E})^T,\quad \boldsymbol{\chi}_{E}=(\chi_{1;E},\chi_{2;E})^{T},
\end{equation}
then we can define $\mathbf{\Phi}_E $ as 
\begin{eqnarray}
\boldsymbol{\Phi}_{E}&=&\mathcal{U}\left((1,0)^T\otimes\boldsymbol{\xi}_{E}+(0,1)^T\otimes\boldsymbol{\chi}_{E}\right)=\mathcal{U}\begin{pmatrix}
\boldsymbol{\xi}_{E} \\
\boldsymbol{\chi}_{E}
\end{pmatrix}\\&=&
\mathcal{U}(\xi_{1;E},\xi_{2;E},\chi_{1;E},\chi_{2;E})^{T} \, ,
\end{eqnarray} 
that satisfies $$(H-E)\mathbf{\Phi}_E=0.$$

The operator $H_{dis}$ in (\ref{Hsoc}) is reducible provided that
\begin{equation}
V'=-V,\quad W^+=-e^{-2i\alpha}(W^-)^{\dagger},\quad W_B=e^{-i\alpha}c_B,\quad W_A=e^{-i\alpha}c_A,
\label{Hdisredassumption}
\end{equation}
with $c_A\equiv c_{A}(x)$ and $c_B\equiv c_{B}(x)$ real-valued functions. Indeed, if we define the unitary matrix $\mathcal{U}\equiv\mathcal{U}_{dis}$ as
\begin{equation}
\mathcal{U}_{dis}=
\frac{\sqrt{2}}{2}\begin{pmatrix}
0 & 1 & 0 & -e^{-i\alpha}\\ 
1 & 0 & -e^{-i\alpha} & 0 \\ 
0 & - e^{i\alpha} & 0 & -1 \\ 
e^{i\alpha}& 0 & 1  & 0
\end{pmatrix}
\, , 
\label{V}
\end{equation}
then there holds
\begin{equation}\label{Hdisred}
\mathcal{U}_{dis}^{-1}H_{dis}\,\mathcal{U}_{dis}=\left(\mathbb{S}_{1}\otimes \mathbf{h_{1}}+\mathbb{S}_{2}\otimes \mathbf{h_{2}}\right),
\end{equation}
where
\begin{equation}
\mathbf{v_{1}}=
\begin{pmatrix}
c_B+V_B & V^\dagger-e^{i\alpha} W^-  \\
V-e^{-i\alpha} (W^-)^\dagger & -c_A+V_A 
\end{pmatrix}
\, , 
\quad 
\mathbf{v_{2}}=
\begin{pmatrix}
-c_B+V_B & V^\dagger+e^{i\alpha} W^-  \\
V+e^{-i\alpha} (W^-)^\dagger & c_A+V_A 
\end{pmatrix}.
\end{equation}
Hermiticity of $\mathbf{v_1}$ and $\mathbf{v_2}$ follows from (\ref{Hdisredassumption}).
It is worth noticing that the components of $\mathbf{v_1}$ and $\mathbf{v_2}$ are linearly independent. It means that for any hermitian $\mathbf{h_1}$ and $\mathbf{h_2}$, we can revert (\ref{H-blocks}) and construct $H_{dis}$  whose components of the potential matrix satisfy (\ref{Hdisredassumption}).

The Hamiltonian $H_{soc}$ is reducible without any additional requirements. Fixing the unitary transformation $\mathcal{U}\equiv\mathcal{U}_{soc}$ as
\begin{equation}
\mathcal{U}_{soc}=
\frac{\sqrt{2}}{2}\begin{pmatrix}
1 & 0 & -e^{-i\alpha} & 0\\ 
0 & 1 & 0 & -e^{-i\alpha} \\ 
0 &  e^{i\alpha} & 0 & 1 \\ 
e^{i\alpha}& 0 & 1  & 0
\end{pmatrix}
\, , \quad \alpha=\pi/2,
\label{V2}
\end{equation}
then we get
\begin{equation}\label{Hsocred}
\mathcal{U}_{soc}^{-1}H_{soc}\,\mathcal{U}_{soc}=\left(\mathbb{S}_{1}\otimes \mathbf{h_{1}}+\mathbb{S}_{2}\otimes \mathbf{h_{2}}\right),
\end{equation}
where
\begin{equation}
\mathbf{v_{1}}=
\begin{pmatrix}
V+\Delta & 0  \\
0 & V-\Delta+\lambda 
\end{pmatrix}
\, , 
\quad 
\mathbf{v_{2}}=
\begin{pmatrix}
V+\Delta &0 \\
0 & V-\Delta-\lambda 
\end{pmatrix}.\label{v1v2Hsoc}
\end{equation}
We can construct the spin-orbit interaction Hamiltonian $H_{soc}$ as $\mathcal{U}_{soc}\left(\mathbb{S}_{1}\otimes\mathbf{h_{1}}+\mathbb{S}_{2}\otimes \mathbf{h_{2}}\right)\mathcal{U}_{soc}^{-1}$ where  $\mathbf{v_1}$ and $\mathbf{v_2}$ are as in (\ref{v1v2Hsoc}). 

Dealing with a $2\times 2$ Darboux transformation is much simpler than the $4\times 4$ transformation, as it is easier to guarantee the hermiticity of the new Hamiltonian while preserving the required form \cite{samsonovstac}. We can use the latter to generate new Hamiltonians
\begin{eqnarray}
\mathbf{\widetilde{h}}_j&=&-i\sigma_1\partial_x+\mathbf{\widetilde{v}_j}=-i\sigma_1\partial_x+\mathbf{v_j}+[\sigma_j,\partial_xU_j\,U_j^{-1}]\\&=&-i\sigma_1\partial_x+\begin{pmatrix}
\widetilde{v}_{j} & \widetilde{a}_{j} \\
\widetilde{a}_{j}^{*} & \widetilde{w}_{j}
\end{pmatrix},\quad j=1,2.\label{widetilde{v_a}}
\end{eqnarray}
such that
\begin{equation}\label{reducedinter}
\mathcal{L}_j\mathbf{h_j}=\mathbf{\widetilde{h}_j}\mathcal{L}_j, \quad \mathcal{L}_j=U_j\partial_xU_j^{-1},\quad \partial_x(U_j^{-1}h_jU_j)=0,\quad j=1,2.
\end{equation}
In this form, we can directly compose the Hamiltonian $\widetilde{H}_{dis}$ as  
\begin{eqnarray}
\widetilde{H}_{dis}&=&\mathcal{U}_{dis}\left(\mathbb{S}_{1}\otimes \mathbf{\widetilde{h}_{1}}+\mathbb{S}_{2}\otimes \mathbf{\widetilde{h}_{2}}\right)\mathcal{U}_{dis}^{-1}\\&=&-i\sigma_3\otimes\sigma_1\partial_x+
\left(\begin{array}{cccc}
\widetilde{V}_A&\widetilde{V}&\widetilde{W}_A&\widetilde{W}^{+}\\
\widetilde{V}^*& \widetilde{V}_B&\widetilde{W}^{-}&\widetilde{W}_B\\
\widetilde{W}_{A}^{*} & (\widetilde{W}^{-})^{*} &\widetilde{V}_A & \widetilde{V}'\\
(\widetilde{W}^{+})^{*} & \widetilde{W}_B^{*} & \widetilde{V}'^{*}&\widetilde{V}_B
\end{array}\right),\label{widetildeVdis}
\end{eqnarray}
where the components of potential term are given explicitly as
\begin{align}
&\widetilde{V}_{A}:=\frac{\widetilde{w}_{1}+\widetilde{w}_{2}}{2} \, , \quad\widetilde{V}_{B}:=\frac{\widetilde{v}_{1}+\widetilde{v}_{2}}{2} \,,\quad \widetilde{V}=-\widetilde{V}'=\frac{\widetilde{a}_{1}^{*}+\widetilde{a}_{2}^{*}}{2} \,, \quad \widetilde{W}^{-}:=\frac{e^{-i\alpha}}{2}\left(\widetilde{a}_{1}-\widetilde{a}_{2} \right),
&& \\
&\widetilde{W}_{A}:=\frac{e^{-i\alpha}}{2}\left(\widetilde{w}_{1}-\widetilde{w}_{2}\right) \, , \quad W_{B}:=-\frac{e^{-i\alpha}}{2}\left(\widetilde{v}_{1}-\widetilde{v}_{2} \right) \, ,\quad \widetilde{W}^{+}:=-\frac{e^{-i\alpha}}{2}\left(\widetilde{a}_{1}^{*}-\widetilde{a}_{2}^*\right) \, .
\end{align}

On the other hand, the construction of $\widetilde{H}_{soc}$ from $\mathbf{\widetilde{h}_j}$ is less straightforward. As we observed in (\ref{v1v2Hsoc}), the potential terms of $\mathbf{\widetilde{h}_j}$ associated with $\widetilde{H}_{soc}$ have to be diagonal matrices. Therefore, we need to generate $\mathbf{\widetilde{v}_j}$ in (\ref{widetilde{v_a}}) with $\widetilde{a}_j=0$, $j=1,2$.
Depending on the properties of the initial Hamiltonian, we can get $\widetilde{H}_{soc}$ by a convenient choice of the matrix $U$. 
Let us fix the initial operator $\mathbf{h_1}$ as 
\begin{equation}
\mathbf{h_1}=-i\sigma_1\partial_x+\omega_3\sigma_3\, ,\quad \omega_3(x)\in\mathbb{R}.
 \label{Samsonov}
 \end{equation} 
The Hamiltonian anticommutes with $\sigma_2$, $\{\mathbf{h_1},\sigma_2\}=0$. We compose the matrix $U_1=(\boldsymbol{\xi}_1,\sigma_2\boldsymbol{\xi}_1)$ from an eigenstate $\boldsymbol{\xi}_1$ of $\mathbf{h_1}$, $(\mathbf{h_1}- \epsilon_1)\boldsymbol{\xi}_1=0$. Then the new Hamiltonian reads as \cite{samsonovstac}
\begin{equation}
\mathbf{\widetilde{h}_1}=-i\sigma_1\partial_x+\widetilde{\omega}_3\sigma_3,\quad \widetilde{\omega}_3=\left(\omega_3+2\frac{\boldsymbol{\xi}_1^T\, \sigma_2\, (\partial_x\,\boldsymbol{\xi}_1)}{\boldsymbol{\xi}_1^T\boldsymbol{\xi}_1}\right).\label{Samsonov2}
\end{equation}
The potential term $\mathbf{\widetilde{v}_1}=\widetilde{\omega}_3\sigma_3$ has the required diagonal form now. We can follow the same steps to get $\mathbf{\tilde{h}_2}$ with a diagonal potential term. Nevertheless, it is unclear how to get $\mathbf{\widetilde{v}_1}\neq\mathbf{\widetilde{v}_2}$  such that $(\mathbf{\widetilde{v}_1})_{11}=(\mathbf{\widetilde{v}_2})_{11}$, which is also required in (\ref{v1v2Hsoc}). Therefore, we can implement the Darboux transformation only partially to get $\mathbf{\widetilde{h_1}}$ whereas we can fix $\mathbf{h_2}$ as 
\begin{equation}
\mathbf{h_2}=-i\sigma_1\partial_x+\widetilde{\omega}_3\,\sigma_0+2(\lambda(x)-\widetilde{\omega}_3)\mathbb{S}_2
\label{h2ad-hoc}
\end{equation} 
such that  (\ref{v1v2Hsoc}) is satisfied by construction. Then we can construct $\widetilde{H}_{soc}$ as
\begin{equation}
\widetilde{H}_{soc}=-i\sigma_0\otimes\sigma_1\,\partial_x+\widetilde{V}_{soc}=\mathcal{U}_{soc}\left(\mathbb{S}_{1}\otimes \mathbf{\widetilde{h}_{1}}+\mathbb{S}_{2}\otimes \mathbf{h_{2}}\right)\mathcal{U}_{soc}^{-1},\quad 
\label{widetildeHsoc}
\end{equation}
where  $(\widetilde{H}_{soc}-E)\widetilde{\mathbf{\Phi}}=0$ is (at least) partially solvable as we suppose that the eigenstates of $\mathbf{\widetilde{h}_1}$ are known, but ${\mathbf{h}}_{2}$ and its eigenstates are still unknown. 

The potential $\widetilde{V}_{soc}$ reads as,
\begin{equation}\widetilde{V}_{soc}=\left(\begin{array}{cccc}\widetilde{\omega}_3&0&0&0\\
0&\lambda-\widetilde{\omega}_3&-i{\lambda}&0\\
0&i{\lambda}&\lambda-\widetilde{\omega}_3&0\\
0&0&0&\widetilde{\omega}_3
\end{array}\right).
\end{equation}
Comparing with (\ref{Hsoc}), we can conclude that the Rashba term is proportional to $\lambda$, whereas electrostatic field $\widetilde{V}$ and the intrinsic spin-orbit coupling term $\widetilde{\Delta}$ are
\begin{equation}
\widetilde{V}=\frac{1}{2}\lambda,\quad  \widetilde{\Delta}=\widetilde{\omega}_3-\frac{\lambda}{2}.
\end{equation}
We will discuss this situation later in the text.

Let us discuss the correspondence between $2\times 2$ Darboux transformation of reduced systems and $4\times 4$ Darboux transformation of the Hamiltonians $\widetilde{H}_{dis}$ and $\widetilde{H}_{soc}$. First, we focus on the case of $\widetilde{H}_{dis}$.
The reduced hamiltonians satisfy the  intertwining relations (\ref{reducedinter}). The operators $H_{dis}$ and $\widetilde{H}_{dis}$ are given by (\ref{Hdisred}) and (\ref{widetildeVdis}).
We can construct the following operator 
\begin{equation}
\mathcal{L}_{dis}=\mathbb{I}_{4\times 4}\partial_{x}-(\partial_{x}U)U^{-1}=\mathcal{U}_{dis}(\mathbb{S}_{1}\otimes \mathcal{L}_{1}+\mathbb{S}_{2}\otimes \mathcal{L}_{2} )\mathcal{U}_{dis}^{-1}\, , 
\label{red-Ldis}
\end{equation}
where the matrix $4\times4$ matrix $U$ can be written as
\begin{equation}
U=\mathcal{U}_{dis}(\mathbb{S}_{1}\otimes U_{1} + \mathbb{S}_{2}\otimes U_{2})\mathcal{U}_{dis}^{-1}\, .
\label{U4x4-red}
\end{equation}
Then, using the orthogonality property $\mathbb{S}_{j}\mathbb{S}_{k}=\delta_{j,k}\mathbb{S}_{j}$ together with  (\ref{reducedinter}), (\ref{Hdisred}) and (\ref{widetildeVdis}), one can check that that there holds
\begin{equation}
\mathcal{L}_{dis}H_{dis}=
\widetilde{H}_{dis}\mathcal{L}_{dis} \, .
\label{LHHLdis}
\end{equation}
Now, the case of $H_{soc}$ and $\widetilde{H}_{soc}$ can be treated in similar manner. The intertwining operator for the partial Darboux transformation can be defined as
\begin{equation}\label{red-Lsoc}
\mathcal{L}_{soc}=\mathcal{U}_{soc}(\mathbb{S}_{1}\otimes \mathcal{L}_{1}+\mathbb{S}_{2}\otimes \sigma_0 )\mathcal{U}_{soc}^{-1}\,.
\end{equation}
Notice that $\mathcal{L}_2$ has been replaced by the identity operator $\sigma_0$. The operator satisfies
\begin{equation}
\mathcal{L}_{soc}H_{soc}=
\widetilde{H}_{soc}\mathcal{L}_{soc} \, .
\end{equation}
It has to be kept in mind that in this case, we require $(\mathbf{\widetilde{v}_1})_{11}=(\mathbf{{v}_2})_{11}$.

A few comments are in order. It is worth noticing that $\mathcal{L}_{dis}$ is a special case of $4\times4$ Darboux transformation (\ref{L}). In contrary, the intertwining operator  $\mathcal{L}_{soc}$ lies outside this class as it cannot be written as $U\partial_xU^{-1}$ for a $4\times4$ matrix $U$.  
When the Darboux transformation $\mathcal{L}$ of a reducible Hamiltonian (\ref{H-blocks}) is reducible, i.e. it can be written like in  (\ref{red-Ldis}) or (\ref{red-Lsoc}), then the new operator $\widetilde{H}$ is also reducible. When $\mathcal{L}$ cannot be transformed into the corresponding block diagonal form by the unitary transformation, reducibility of $\widetilde{H}$ is not granted. We will discuss the non-reducible Darboux transformations later in the text.

So far, we have shown that the reduction scheme can be used for construction of a solvable $4\times4$ Hamiltonian that possesses the required form. We found that whenever we get hermitian $2\times2$ hamiltonians $\mathbf{\widetilde{h}_1}$ and $\mathbf{\widetilde{h}_2}$ as in (\ref{widetilde{v_a}}), we can construct $\widetilde{H}_{dis}$ with the potential as in (\ref{widetildeVdis}). In this case, the reducible Darboux transformation $\mathcal{L}_{dis}$ defined in (\ref{red-Ldis}) is {\it form-preserving}, see (\ref{LHHLdis}). The Hamiltonian $\widetilde{H}_{soc}$ with spin-orbit interaction can be constructed from two $2\times2$ operators whose potentials satisfy (\ref{widetildeHsoc}). Here, Darboux transformation is only partial as we use it to get $\mathbf{\widetilde{h}_1}$. The operator $\mathbf{h_2}$ has to be fixed such that $(\mathbf{\widetilde{v}_1})_{11}=(\mathbf{{v}_2})_{11}$. 
Finally, let us notice that although we focused on the specific operators $H_{dis}$ and $H_{soc}$, the results are applicable for any reducible Hamiltonian that satisfies (\ref{H-blocks}).

\section{$2\times 2$ reflectionless models via Darboux transformation }
\label{sec:2x2-reflectionless}
We focus on the construction of new exactly solvable models based on the free particle system described by $2\times2$ Hamiltonian. Depending on the explicit form of the Darboux transformation, the new systems may inherit physical characteristics of the original system, such as reflection-free scattering. In this section, we will discuss construction and main properties of these systems.

We consider a matrix Hamiltonian of dimension $2\times 2$ with a constant matrix potential; that is, its matrix elements are independent of the position and time coordinates,
\begin{equation}
\mathbf{h}=-i
\sigma_1
\partial_{x}+
\begin{pmatrix}
v& a \\
a^{*} & w
\end{pmatrix}
\, , \quad v,w\in\mathbb{R} \, , \quad a\in\mathbb{C} \, .
\label{h}
\end{equation}
The eigenvalue equation 
\begin{equation}
h\boldsymbol{\psi}_{E}=E\boldsymbol{\psi}_{E} \, , 
\label{hstac}
\end{equation}
has the following fundamental set of solutions,
\begin{eqnarray}
\boldsymbol{\psi}_{E}&=& 
\begin{pmatrix}
\psi_{1;E} \\
\psi_{2;E}
\end{pmatrix} =
e^{-i(\operatorname{Re}a)x} \begin{pmatrix}
\cosh(\kappa_{E}x) \\
\frac{i}{w-E}\left( \operatorname{Im}a\cosh(\kappa_{E}x)+\kappa_{E}\sinh(\kappa_{E}x) \right)
\end{pmatrix},\nonumber\\
\overline{\boldsymbol{\psi}}_{E}&=&
\begin{pmatrix}
\overline{\psi}_{1;E} \\
\overline{\psi}_{2;E}
\end{pmatrix}  =
e^{-i(\operatorname{Re}a)x} \begin{pmatrix}
\frac{i}{v-E}\left( -(\operatorname{Im}a)\cosh(\kappa_{E}x)+\kappa_{E}\sinh(\kappa_{E}x) \right) \\
\cosh(\kappa_{E}x) 
\end{pmatrix},
\label{2x2spinor1}
\end{eqnarray}
where
\begin{equation}
\kappa^{2}_{E}=(\operatorname{Im}a)^{2}-(v-E)(w-E) \, .
\end{equation}
The solutions (\ref{2x2spinor1}) have exponential-like behavior for real $\kappa_E$, whereas they are oscillating for complex $\kappa_E$.
The parameter $\kappa_{E}$ is real for $E$ that lies within the following interval,
\begin{equation}
E\in(\epsilon_{-},\epsilon_{+}) \, , \quad \epsilon_{\pm}=\frac{(v+w)\pm\sqrt{(v-w)^2+4(\operatorname{Im}a)^{2}}}{2} \, .
\label{cond1}
\end{equation}
The solutions in~\eqref{2x2spinor1} are not normalizable, independently on the value of $\kappa_E$.
The Hamiltonian $\mathbf{h}$ has translational symmetry, so that the spinors $ {\boldsymbol{\psi}}_{E}(x+\alpha)$ and $\overline{\boldsymbol{\psi}}_{E}(x+\alpha)$, $\alpha\in\mathbb{\mathbb{C}}$, are solutions of (\ref{hstac}) as well. We will use this fact later in the text. 

We can proceed to the construction of the new models with the use of Darboux transformation. As we discussed in section \ref{section3}, the intertwining operator as well as the new Hamiltonian are defined in terms of the matrix $U$ that satisfies $\mathbf{h}\,U=U\,\operatorname{diag}(\epsilon_1,\epsilon_2)$, see (\ref{U}).  We fix $\epsilon_{1}$ and $\epsilon_{2}$ as $\epsilon_{1},\epsilon_{2}\in(\epsilon_{-},\epsilon_{+})$ so that the corresponding eigenstates (\ref{2x2spinor1}) are exponentially expanding for large $|x|$. 
For convenience\footnote{This choice provides more compact formulas. It is also easier to discuss invertibility of the matrix $U$.}, we construct the seed matrix $U$ by combining solutions in the form $\boldsymbol{\psi}_{\epsilon_{1}}$ and $\overline{\boldsymbol{\psi}}_{\epsilon_{2}}$. That is, 
\begin{multline}
U=\left( \boldsymbol{\psi}_{\epsilon_{1}},\overline{\boldsymbol{\psi}}_{\epsilon_{2}} \right)=
\begin{pmatrix}
\psi_{1;\epsilon_{1}} & \overline{\psi}_{1;\epsilon_{2}}\\
\psi_{2;\epsilon_{1}} & \overline{\psi}_{2;\epsilon_{2}}
\end{pmatrix}
=e^{-i(\operatorname{Re}a)x}\times \\
\begin{pmatrix}
\cosh(z_{1}) & \frac{i}{v-\epsilon_{2}}\left( -(\operatorname{Im}a)\cosh({z}_{2})+\kappa_{\epsilon_{2}}\sinh({z}_{2}) \right) \\
\frac{i}{w-\epsilon_{1}}\left( (\operatorname{Im}a)\cosh(z_{1})+\kappa_{\epsilon_{1}}\sinh(z_{1}) \right) & \cosh({z}_{2}) 
\end{pmatrix}
\, ,
\label{U1}
\end{multline}
with $\kappa_{\epsilon_{j}}=\kappa_{E\rightarrow\epsilon_{j}}$, for $j=1,2$, and
\begin{equation}
z_{1}\equiv z_{1}(x)=\kappa_{\epsilon_{1}}x+\delta_{1} \, , \quad {z}_{2}\equiv z_{2}(x)=\kappa_{\epsilon_{2}}x+{\delta}_{2} \, .
\label{zz}
\end{equation}
The determinant of $U$ takes the explicit form
\begin{equation}
\operatorname{det}(U)= \frac{e^{-2i(\operatorname{Re}a)x}\cosh(z_{1})\cosh({z}_{2})}{(v-\epsilon_{2})(w-\epsilon_{1})}\mathcal{D}(x) \, ,
\label{detU1}
\end{equation}
where
\begin{equation}
\mathcal{D}(x)=(v-\epsilon_{2})(w-\epsilon_{1})+\left( \kappa_{\epsilon_{1}}\tanh(z_{1})+\operatorname{Im}a\right)\left( \kappa_{\epsilon_{2}}\tanh({z}_{2})-\operatorname{Im}a\right) .
\label{detU2}
\end{equation}
From~\eqref{detU1}-\eqref{detU2}, it is clear that the zeros of $\operatorname{det}(U)$ correspond to those of the function $\mathcal{D}(x)$. By fixing $v<w$, $\operatorname{Im}a>0$ and  $\epsilon_1,\epsilon_2\in(v,w)$, it is possible to analytically show that  $\mathcal{D}(x)$ is a nodeless function if the following condition holds\footnote{This condition is sufficient but not necessary. Numerical analysis confirms that there is a wide range of parameters where the condition is not satisfied but $\mathcal{D}(x)$ is still node-less. A similar condition could be found for $w<v$, $\operatorname{Im}a<0$ and  $\epsilon_1,\epsilon_2\in(w,v)$.}
\begin{equation}
(w-\epsilon_{1})(\epsilon_{2}-v)+\operatorname{Im}a^2>\kappa_{\epsilon_{1}}\kappa_{\epsilon_{2}}+(\operatorname{Im}a)(\kappa_{\epsilon_{1}}+\kappa_{\epsilon_{2}}) \, .
\label{cond2}
\end{equation}

\begin{figure}
\centering
\subfloat[][$\widetilde{w}$]{\includegraphics[width=0.4\textwidth]{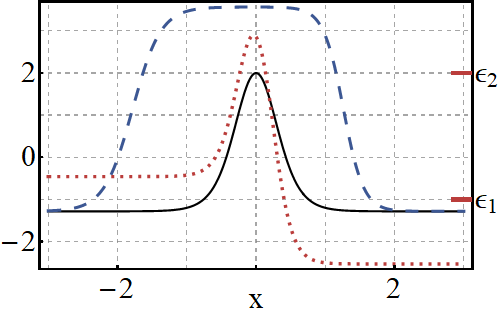}}
\hspace{2mm}
\subfloat[][$\operatorname{Im}(\widetilde{a}^{*}(x))$]{\includegraphics[width=0.4\textwidth]{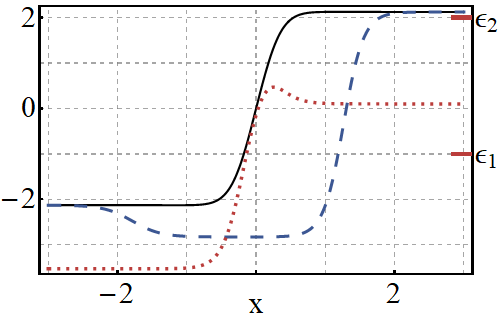}}
\caption{Components $\widetilde{\omega}(x)$ (a) and $\operatorname{Im}(\widetilde{a}^{*}(x))$ introduced in~\eqref{VV1}-\eqref{VV2}. In both cases, we have considered factorization energies $\epsilon_{1}=-1$ and $\epsilon=2$, together with $v=-2$ and $w=5$. Moreover, the following set of parameters were used: $\{\delta_{1}=0,{\delta}_{2}=0,A=0\}$ (black-solid), $\{\delta_{1}=4,{\delta}_{2}=-4,a=0\}$ (blue-dashed), and $\{\delta_{1}=0,{\delta}_{2}=0,a=2i\}$ (red-dotted).}
\label{fig:pot-2x2-1}
\end{figure}

From all the previous considerations, we can implement the Darboux transformation of Sec.~\ref{sec:Dirac},
\begin{equation}\label{reflectionlessL}
\mathcal{L}=U\partial_xU^{-1},
\end{equation}
and construct the new Hamiltonian
\begin{equation}
\mathbf{\widetilde{h}}=\mathbf{h}-[U'U^{-1},\gamma] =-i\sigma_{1}\partial_{x}+\mathbf{\widetilde{v}}(x) 
=-i\sigma_{1}\partial_{x}+\begin{pmatrix}
\widetilde{v} & \widetilde{a} \\
\widetilde{a}^{*} & \widetilde{w}
\end{pmatrix}\, ,
\label{VV1}
\end{equation}
where the components of the potential are 
\begin{eqnarray}
\widetilde{v}&=&w+\epsilon_{2}-\epsilon_{1}-2(\epsilon_{1}-\epsilon_{2})\frac{(w-\epsilon_{1})(v-\epsilon_{2})}{\mathcal{D}(x)},\nonumber\\
\widetilde{w}&=&v-\epsilon_{2}+\epsilon_{1}+2(\epsilon_{1}-\epsilon_{2})\frac{(w-\epsilon_{1})(v-\epsilon_{2})}{\mathcal{D}(x)},\nonumber\\
\widetilde{a}&=&\operatorname{Im}a+(\epsilon_{1}-\epsilon_{2})\frac{(\operatorname{Im}a)(v-w+\epsilon_{1}-\epsilon_{2})+(v-\epsilon_{2})\kappa_{\epsilon_{1}}\tanh(z_1)+(w-\epsilon_{1})\kappa_{\epsilon_{2}}\tanh({z}_{2})}{\mathcal{D}(x)}.\nonumber\\\label{widetildeVWA}
\label{VV2}
\end{eqnarray}
Here, $z_1$ and $z_2$ are linear functions of $x$, see (\ref{zz}), whereas $\mathcal{D}(x)$ is defined in (\ref{detU1}). 
To illustrate the properties of $\mathbf{\widetilde{h}}$, in Fig.~\ref{fig:pot-2x2-1} we depict the diagonal component $\widetilde{w}$ and the anti-diagonal component $\operatorname{Im}\widetilde{a}$.

\begin{figure}
\centering
\subfloat[][]{\includegraphics[width=0.3\textwidth]{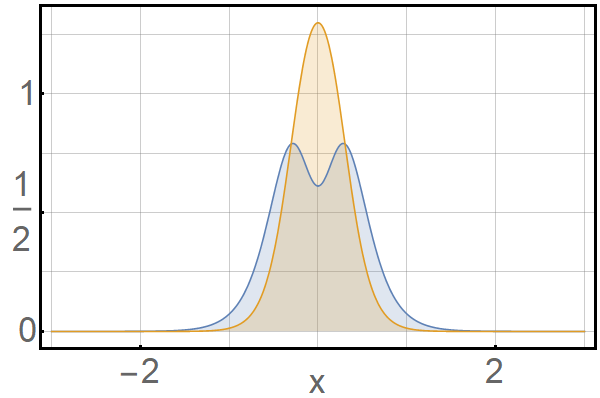}
\label{fig:ms2x2-1-a}}
\hspace{2mm}
\subfloat[][]{\includegraphics[width=0.3\textwidth]{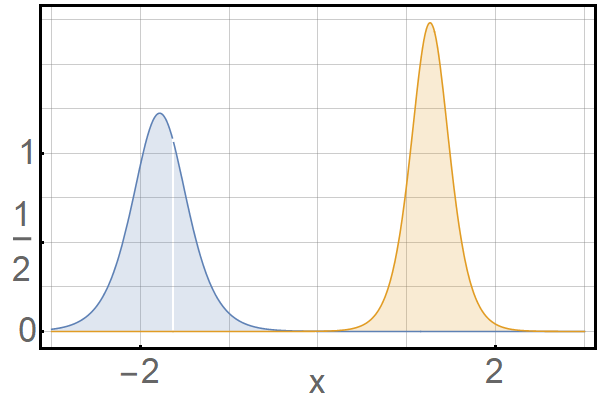}
\label{fig:ms2x2-1-b}}
\hspace{2mm}
\subfloat[][]{\includegraphics[width=0.3\textwidth]{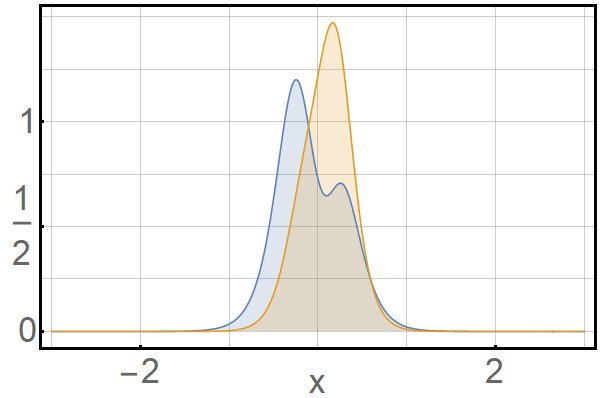}
\label{fig:ms2x2-1-c}}
\caption{Probability densities $\mathcal{P}_{\epsilon_{1}}(x)$ (blue) and $\mathcal{P}_{\epsilon_{2}}$ (orange) associated to the missing states in~\eqref{missing-2x2}. For all the cases, we have used the factorization energies $\epsilon_{1}=-1$ and $\epsilon_{2}=2$, together with $v=-2$ and $w=5$. The rest of parameter have been selected as $\{\delta_{1}=0,{\delta}_{2}=0,a=0\}$ (a), $\{\delta_{1}=4,{\delta}_{2}=-4,a=0\}$ (b), and $\{\delta_{1}=0,{\delta}_{2}=0,a=2i\}$ (c).}
\label{fig:ms2x2-1}
\end{figure}

Contrary to the initial free particle Hamiltonian $\mathbf{h}$, the operator $\mathbf{\widetilde{h}}$ admits two bound states $\widetilde{\boldsymbol{\psi}}_{\epsilon_{1}}$ and $\widetilde{\boldsymbol{\psi}}_{\epsilon_{2}}$ associated with the energies $\epsilon_{1}$ and $\epsilon_{2}$, respectively, see (\ref{missing}). After some calculation, we get
\begin{equation}
\begin{aligned}
&\widetilde{\boldsymbol{\psi}}_{\epsilon_{1}}=\frac{e^{-i(\operatorname{Re}a)x}\operatorname{sech}(z_{1})}{\mathcal{N}_{\epsilon_{1}}\mathcal{D}(x)}
\begin{pmatrix}
1 \\
\frac{i}{(v-\epsilon_{2})}\left(-\operatorname{Im}a+\kappa_{\epsilon_{2}}\tanh({z}_{2})\right)
\end{pmatrix}
\, , \\
&\widetilde{\boldsymbol{\psi}}_{\epsilon_{2}}=\frac{e^{-i(\operatorname{Re}a)x}\operatorname{sech}({z}_{2})}{\mathcal{N}_{\epsilon_{2}}\mathcal{D}(x)}
\begin{pmatrix}
\frac{i}{(w-\epsilon_{1})}\left(\operatorname{Im}a+\kappa_{\epsilon_{1}}\tanh(z_{1})\right) \\
1
\end{pmatrix}
\, ,
\end{aligned}
\label{missing-2x2}
\end{equation}
with $\mathcal{N}_{\epsilon_{j}}$ the corresponding normalization factors. As long as $\epsilon_1\neq\epsilon_2$ or $\kappa_{\epsilon_j}\neq0$, both $\widetilde{\boldsymbol{\psi}}_{\epsilon_{1}}$ and $\widetilde{\boldsymbol{\psi}}_{\epsilon_{2}}$ vanish exponentially for large $\vert x\vert$. Since $\mathcal{D}(x)\neq 0$ on the whole real line, both spinors in~\eqref{missing-2x2} have finite norm. In turn, if $\epsilon_{1}=\epsilon_{2}=\epsilon\in(\epsilon_{-},\epsilon_{+})$, the potential $\mathbf{\widetilde{v}}$ corresponds to a constant matrix so that none of the missing states (\ref{missing-2x2}) are square integrable.
On the other hand, for $\epsilon_{1}=\epsilon_{+}$ (or $\epsilon_{1}=\epsilon_{-}$) and $\epsilon_2\neq\epsilon_{\pm}$, we have $\kappa_{\epsilon_{1}}=0$ and $\widetilde{\boldsymbol{\psi}}_{\epsilon_{1}}$ does not have finite norm. We thus only get $\widetilde{\boldsymbol{\psi}}_{\epsilon_{2}}$ as the bound state solution. Analogous results are obtained for $\epsilon_{2}=\epsilon_{\pm}$.
The probability density $\mathcal{P}_{\epsilon_{j}}(x)$ associated to the missing eigensolutions is shown in Fig.~\ref{fig:ms2x2-1} for several values of the respective parameters. From the latter, it can be noticed that the probability density becomes a symmetric function for $\operatorname{Im}a=\delta_{1}=\delta_{2}=0$. The presence of the phases $\delta_{k}$ and $\operatorname{Im}a$ induces an asymmetry on the probability densities, as depicted in Figs.~\eqref{fig:ms2x2-1-b}-\eqref{fig:ms2x2-1-c}. 

The intertwining operator $\mathcal{L}$ provides a one-to-one mapping of the scattering states\footnote{$\mathcal{L}$ annihilates the two eigenstates from the interior of the energy band.}. It can serve well in the analysis of the scattering properties of the new system. For illustration, let us consider  is an incoming plane wave solution $\boldsymbol{\psi_E}$,
\begin{equation}
\mathbf{h}\,\boldsymbol{\psi_E}=E\boldsymbol{\psi_E},\quad \boldsymbol{\psi_E}=e^{\kappa_E x}\left(1,\frac{i(E-v)}{a+\kappa_E}\right)^T ,\quad E\not\in (\epsilon_-,\epsilon_+) .
\end{equation}
The intertwining operator $\mathcal{L}$ transforms the scattering state into spinor $\boldsymbol{\widetilde{\psi}_E}=\mathcal{L}\,\boldsymbol{\psi_E}$ such that
\begin{equation}
\mathbf{\widetilde{h}}\,\boldsymbol{\widetilde{\psi}_E}=E\boldsymbol{\widetilde{\psi}_E}.
\end{equation}
The potential term of $\mathbf{\widetilde{h}}$ tends to a constant matrix for large $|x|$. Therefore, it can be expected that the scattering states can be identified with plane waves for $x\rightarrow\pm\infty$. This is indeed the case. There holds $\lim_{x\rightarrow\pm\infty}U_xU^{-1}=w_{\pm}$ where $w_{\pm}$ are constant $4\times4$ matrices whose explicit form is calculated in the Appendix.
The operator $\mathcal{L}$ acts asymptotically as 
\begin{equation}
\mathcal{L}\,\boldsymbol{\psi_E}\rightarrow (\partial_x+w_{\pm})\boldsymbol{\psi_E}=e^{\kappa_E x}(\kappa_E\sigma_0+w_{\pm})\left(1,\frac{i(E-v)}{a-i\kappa_E}\right)^T,\quad x\rightarrow\pm\infty.
\end{equation}  
The action of the operator $\mathcal{L}$ on the scattering state $\boldsymbol{\psi_E}$ does not change its momentum. As there is no back-scattered wave, the potential barrier represented by $\mathbf{\widetilde{v}}$ in (\ref{VV1}) is \textit{reflectionless} for any energy $E\not\in (\epsilon_-,\epsilon_+).$
We can also write
\begin{equation}
\mathcal{L}\,\boldsymbol{\psi_E}\vert_{x\rightarrow+\infty}=(\kappa_E\sigma_0+w_{+})(\kappa_E\sigma_0+w_{-})^{-1}\mathcal{L}\,\boldsymbol{\psi_E}\vert_{x\rightarrow-\infty}.
\end{equation}

It is worth noticing that Darboux transformation of the free-particle system described by $2\times2$ Dirac Hamiltonian was discussed in \cite{Cor14}. The Hamiltonian (\ref{h}), in comparison to the one used in \cite{Cor14}, is more general and lacks the chiral symmetry. Therefore our current results extend those of \cite{Cor14}. 


\section{Reflectionless models of distortion scattering and spin-orbit interaction\label{section5}}

In this section, we construct solvable systems described by $\widetilde{H}_{dis}$ or $\widetilde{H}_{soc}$ with the use of (\ref{widetildeVdis}) or (\ref{widetildeHsoc}), where we identify $\mathbf{\widetilde{h}_1}$ and $\mathbf{\widetilde{h}_2}$ with the reflectionless Hamiltonians discussed in the previous section. Let us start with the case of the Hamiltoinan for distortion scattering.

\subsection{Reflectionless distortion Hamiltonian\label{Hsocreflectionless}}
Let us recall that the reducible Hamiltonian  $\widetilde{H}_{dis}$ can be written in terms of two $2\times2$ Dirac operators $\mathbf{\widetilde{h}_1}$ and $\mathbf{\widetilde{h}_2}$, see (\ref{widetildeVdis}),
\begin{equation}
\widetilde{H}_{dis}=\mathcal{U}_{dis}\left(\mathbb{S}_{1}\otimes \mathbf{\widetilde{h}_{1}}+\mathbb{S}_{2}\otimes \mathbf{\widetilde{h}_{2}}\right)\mathcal{U}_{dis}^{-1}.\label{reducible2}
\end{equation}  
We shall identify the reduced operators $\mathbf{\widetilde{h}_j}$, $j=1,2$, with $\mathbf{\widetilde{h}}$ in (\ref{VV1}). The potential term $\mathbf{\widetilde{v}}$ of the later Hamiltonian depends on the parameters $v$, $a$, $w$, $\epsilon_1$ and $\epsilon_2$, $\delta_1$ and $\delta_2$. These parameters can acquire different values in $\mathbf{\widetilde{h}_1}$ and $\mathbf{\widetilde{h}_2}$. We define the two operators in the following manner,
\begin{eqnarray}
\mathbf{\widetilde{h}_{1}}&=&\mathbf{\widetilde{h}}\vert_{v\rightarrow v_1,a\rightarrow a_1,w\rightarrow w_1}=-i\sigma_1\partial_x+\begin{pmatrix}
\widetilde{v}_1 & \widetilde{a}_1 \\
\widetilde{a}_1^{*} & \widetilde{w}_1
\end{pmatrix},\nonumber\\
\mathbf{\widetilde{h}_{2}}&=&\mathbf{\widetilde{h}}\vert_{v\rightarrow v_2,a\rightarrow a_2,w\rightarrow w_2,\epsilon_1\rightarrow\epsilon_3,\epsilon_2\rightarrow\epsilon_4,\delta_1\rightarrow\delta_3,\delta_2\rightarrow\delta_4}\nonumber\\&=&-i\sigma_1\partial_x+\begin{pmatrix}
\widetilde{v}_2 & \widetilde{a}_2 \\
\widetilde{a}_2^{*} & \widetilde{w}_2
\end{pmatrix}.\label{widetildeh1h2}
\end{eqnarray}
We select $\epsilon_1$, $\epsilon_2$, $\epsilon_3$ and $\epsilon_4$ such that $\kappa_{\epsilon_{1(2)}}^{(1)}$ and $\kappa_{\epsilon_{3(4)}}^{(2)}$ are real, where $$\kappa^{(j)}_{E}=\sqrt{|\operatorname{Im}a_j|^2-(v_j-E)(w_j-E)},\quad j=1,2.$$
Then the potential terms of both $\mathbf{\widetilde{h}_1}$ and $\mathbf{\widetilde{h}_2}$ are regular and asymptotically constant, see (\ref{widetildeVWA}) with an appropriate substitution from (\ref{widetildeh1h2}). 

The eigenstates of $\mathbf{\widetilde{h}_1}$ and $\mathbf{\widetilde{h}_2}$ can be obtained from (\ref{hstac}) with the use of the intertwining operator (\ref{reflectionlessL}). The missing states associated with $\mathbf{\widetilde{h}_1}$ and $\mathbf{\widetilde{h}_2}$ can be deduced from (\ref{missing-2x2}). They are
\begin{eqnarray}
&&\widetilde{\boldsymbol{\xi}}_{\epsilon_{1}}=\widetilde{\boldsymbol{\psi}}_{\epsilon_{1}}\vert_{v\rightarrow v_1,a\rightarrow a_1,w\rightarrow w_1},\quad \widetilde{\boldsymbol{\xi}}_{\epsilon_{2}}=\widetilde{\boldsymbol{\psi}}_{\epsilon_{2}}\vert_{v\rightarrow v_1,a\rightarrow a_1,w\rightarrow w_1},
\label{missing1}\end{eqnarray}
and 
\begin{eqnarray}
\widetilde{\boldsymbol{\chi}}_{\epsilon_{3}}&=&\widetilde{\boldsymbol{\psi}}_{\epsilon_{1}}\vert_{v\rightarrow v_2,a\rightarrow a_2,w\rightarrow w_2,\epsilon_1\rightarrow\epsilon_3,\epsilon_2\rightarrow\epsilon_4,\delta_1\rightarrow\delta_3,\delta_2\rightarrow\delta_4},\\ \widetilde{\boldsymbol{\chi}}_{\epsilon_{4}}&=&\widetilde{\boldsymbol{\psi}}_{\epsilon_{2}}\vert_{v\rightarrow v_2,a\rightarrow a_2,w\rightarrow w_2,\epsilon_1\rightarrow\epsilon_3,\epsilon_2\rightarrow\epsilon_4,\delta_1\rightarrow\delta_3,\delta_2\rightarrow\delta_4}.\label{missing2}
\end{eqnarray}
and satisfy
\begin{equation}
\mathbf{\widetilde{h}_1}\widetilde{\boldsymbol{\xi}}_{\epsilon_{1,2}}=\epsilon_{1,2}\widetilde{\boldsymbol{\xi}}_{\epsilon_{1,2}},\quad \mathbf{\widetilde{h}_2}\,\widetilde{\boldsymbol{\chi}}_{\epsilon_{3,4}}=\epsilon_{3,4}\,\widetilde{\boldsymbol{\chi}}_{\epsilon_{3,4}}.
\end{equation}



Now, we can construct the Hamiltonian $\widetilde{H}_{dis}$ by inserting the explicit form of $\mathbf{\widetilde{h}_1}$ and $\mathbf{\widetilde{h}_2}$ from (\ref{widetildeh1h2}) into (\ref{widetildeVdis}). As we argued in the end of the preceeding section, the potential $\widetilde{V}_{dis}$ is reflectionless as it does not induce any backscattering.  It is rather straightforward to get the explicit form of its components. For instance, the first diagonal component $\widetilde{V}_A$ can be obtained as  
\begin{equation}
\widetilde{V}_{A}=\frac{\widetilde{w}\vert_{v\rightarrow v_1,a\rightarrow a_1,w\rightarrow w_1}}{2}+\frac{\widetilde{w}\vert_{v\rightarrow v_2,a\rightarrow a_2,w\rightarrow w_2,\epsilon_1\rightarrow\epsilon_3,\epsilon_2\rightarrow\epsilon_4,\delta_1\rightarrow\delta_3,\delta_2\rightarrow\delta_4}}{2} \,  ,\end{equation}
where $\widetilde{w}$ is given in (\ref{widetildeVWA}). The other elements of $\widetilde{V}_{dis}$ can be obtained in exactly the same manner. Additionally, there hold the following relations between the elements,
\begin{align}
&\widetilde{V}_{B}:=-\widetilde{V}_{A}+\frac{w_{1}+w_{2}+v_{1}+v_{2}}{2} \,  ,\quad
\widetilde{W}_{B}=\widetilde{W}_{A}+\frac{e^{-i\alpha}}{2}(w_{2}+v_{2}-w_{1}-v_{1}) \, ,\\& \widetilde{W}^{-}:=\widetilde{W}^{+}+e^{-i\alpha}\operatorname{Re}(a_{1}-a_{2}) \, .
\end{align}

\begin{figure}
\centering
\subfloat[][]{\includegraphics[width=0.3\textwidth]{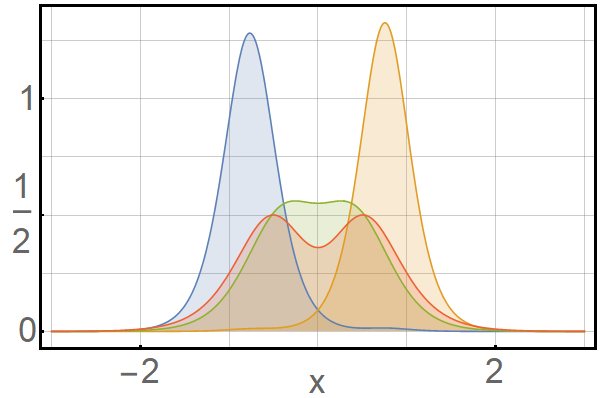}
\label{dist-fig-3}}
\subfloat[][]{\includegraphics[width=0.3\textwidth]{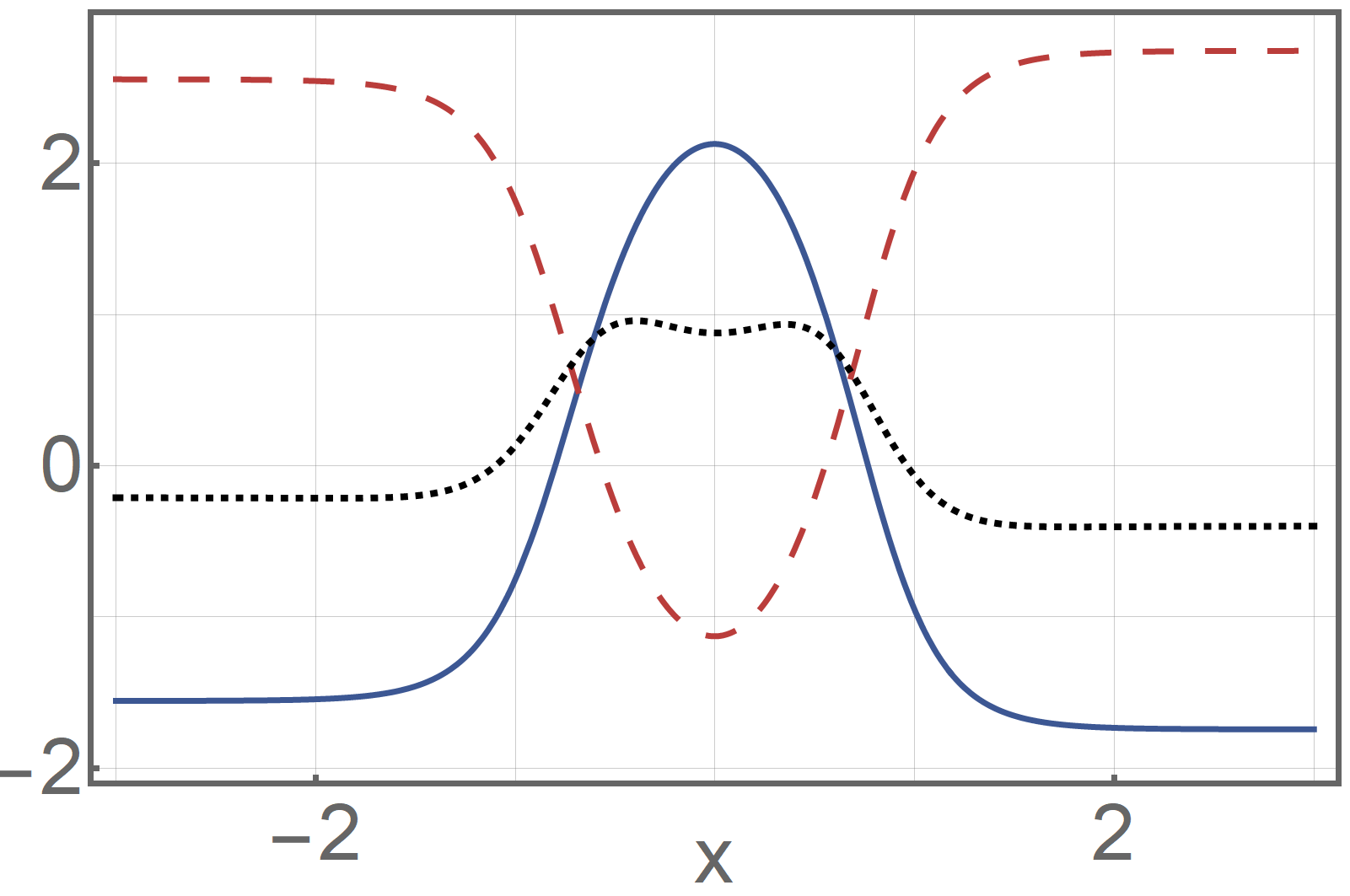}
\label{dist-fig-1}}
\hspace{2mm}
\subfloat[][]{\includegraphics[width=0.3\textwidth]{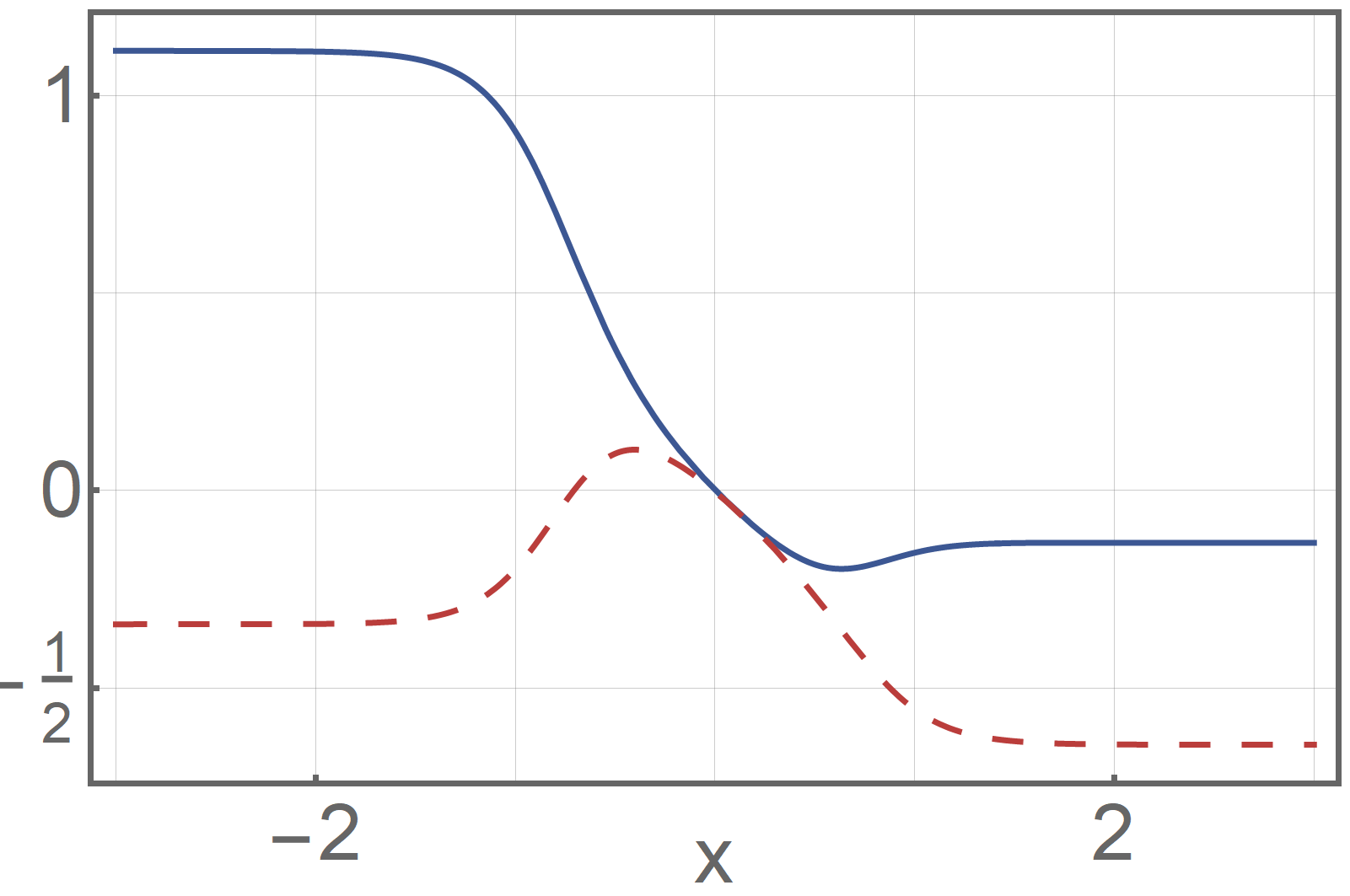}
\label{dist-fig-2}}
\caption{(a) Probability densities $\mathcal{P}_{\epsilon_{j}}(x)$ associated with the missing eigensolutions $\widetilde{\boldsymbol{\Phi}}_{\epsilon_{j}}$ for $\epsilon_{1}$ (blue), $\epsilon_{2}$ (orange), $\epsilon_{3}$ (green), and $\epsilon_{4}$ (red). (b) Matrix potential entries $\widetilde{V}_{A}$ (blue-solid), $\widetilde{V}_{B}$ (red-dashed), and $\widetilde{W}_{A}$(black-dotted), together with (c) $\operatorname{Im}W^{+}$ (blue-solid) and $\operatorname{Im}V$ (red-dashed). In all cases, we have fixed the complex-phase $\alpha=0$ and the remaining parameters, previously introduced in~\eqref{widetildeh1h2}, as $v_{1}=3$, $v_{2}=2.5$, $w_{1}=-2$, $w_{2}=-1.5$, $a_{1}=i$, $a_{2}=1$, $\epsilon_{1}=1.25$, $\epsilon_{2}=0.25$, $\epsilon_{3}=0.75$, $\epsilon_{4}=-0.5$, $\delta_{1}=1$, ${\delta}_{2}=-1$, $\delta_{3}={\delta}_{4}=0$.}
	\label{dist-fig}
\end{figure}

Therefore, we prefer to illustrate their behavior in figure instead of presenting their explicit forms.
In Fig.~\ref{dist-fig} we depict the behavior of the matrix elements of the reflectionless distortion interaction $\widetilde{V}_{dis}$. For simplicity, we have fixed $\alpha=0$. We thus show the shape of the real-valued matrix inputs $\widetilde{V}_{A}$, $\widetilde{V}_{B}$, and $\widetilde{W}_{A}$ in Fig.~\ref{dist-fig-1}, whereas, in Fig.~\ref{dist-fig-2}, we depict the imaginary part of $\widetilde{W}^{+}$ and $\widetilde{V}$. Note that the real part of $\widetilde{W}^{+}$ and $\widetilde{V}$ take constant values since $\alpha=0$. Moreover, $\widetilde{W}_{B}$ and $\widetilde{W}^{-}$ differ from $\widetilde{W}_{A}$ and $\widetilde{W}^{+}$, respectively, by a constant shift. We thus omit $\widetilde{W}_{B}$ and $\widetilde{W}^{-}	$ in Fig.~\ref{dist-fig}.

The distortion Hamiltonian has up to four\footnote{It depends on the actual values of $\epsilon_j$, $j=1,2,3,4,$ see discussion below (\ref{missing-2x2}).} bound states that are based on the missing state of $\mathbf{\widetilde{h}_1}$  and $\mathbf{\widetilde{h}_2}$, see (\ref{missing1}) and (\ref{missing2}), respectively. They can be defined as
\begin{equation}
\widetilde{\boldsymbol{\Phi}}_{\epsilon_{1,2}}=\mathcal{U}_{dis}\left((1,0)^T\otimes\widetilde{\boldsymbol{\xi}}_{\epsilon_{1,2}}\right)\quad \widetilde{\boldsymbol{\Phi}}_{\epsilon_{3,4}}=\mathcal{U}_{dis}\left((0,1)^T\otimes\widetilde{\boldsymbol{\chi}}_{\epsilon_{3,4}}\right).
\label{4x4missing}
\end{equation}
By construction, their probability density  is identical to (\ref{missing1}) or (\ref{missing2}), e.g. $\widetilde{\boldsymbol{\Phi}}_{\epsilon_{1,2}}^\dagger\widetilde{\boldsymbol{\Phi}}_{\epsilon_{1,2}}\sim (\widetilde{\boldsymbol{\xi}}_{\epsilon_{1}})^\dagger \widetilde{\boldsymbol{\xi}}_{\epsilon_{1}}$. Therefore, Fig.~\ref{dist-fig-3} serves well for illustration of density of probability of (\ref{4x4missing}), while Fig.~\ref{dist-fig-1}-\ref{dist-fig-2} depicts the behavior of the elements of $\widetilde{V}_{dis}$.

\subsection{Spin-orbit coupling case}
\label{subsec:spin-orbit}

In this section, we focus on systems with \textit{spin-orbit} interaction described by the corresponding Hamiltonian $H_{soc}$ in~\eqref{Hsoc}. In order to exploit the Darboux transformation, we construct the new Hamiltonian $\widetilde{H}_{soc}$ via (\ref{widetildeHsoc}),
\begin{equation}\nonumber
\widetilde{H}_{soc}=\mathcal{U}_{soc}\left(\mathbb{S}_1\otimes\mathbf{\widetilde{h}_1}+\mathbb{S}_2\otimes \mathbf{h_2}\right)\mathcal{U}_{soc}^{-1}.
\end{equation}
As we discussed in section \ref{section3}, the potential matrices $\mathbf{\widetilde{v}_1}$ and $\mathbf{v_2}$ of $\mathbf{\widetilde{h}_1}$ and $\mathbf{h_2}$, respectively, should be diagonal. Additionally, there should hold that $(\mathbf{\widetilde{v}_1})_{11}=(\mathbf{v_2})_{11}$.
Likewise in the previous subsection, we identify $\mathbf{\widetilde{h}_1}$ with $\mathbf{\widetilde{h}}$ in (\ref{VV1}),
\begin{equation}
\mathbf{\widetilde{h}_1}=\mathbf{\widetilde{h}}\vert_{v\rightarrow v_1,a\rightarrow a_1,w\rightarrow w_1},\quad v_1,w_1,a_1,\delta_1,\delta_2\in\mathbb{R}.
\end{equation}
Additionally, we fix the parameters of the potential term of $\mathbf{\widetilde{h}_1}$ as follows
\begin{align}
 w_{1}=-v_{1} \, , \quad \delta_{1}=\delta_{2}=0 \, , \quad a_{1}=0 \, ,\quad \epsilon_{2}=-\epsilon_{1},\quad \epsilon_1\in(0,v_1) \, ,\quad v_1>0.
\label{soc-parameters}
\end{align}
Then the operator $\mathbf{\widetilde{h}_1}$ acquires the form
\begin{equation}
\mathbf{\widetilde{h}_1}=-i\sigma_1\partial_x+\widetilde{v}_{1}\sigma_3 \, ,\quad \widetilde{v}_{1}(x)=v_1-\frac{2\kappa_{\epsilon_1}^2}{v_1+\epsilon_1\cosh \kappa_{\epsilon_1}x} \, ,
\label{soc-matrix-entries}
\end{equation}
where $\kappa_{\epsilon_1}=\sqrt{v_1^2-\epsilon_1^2}$.
Let us notice that for this specific choice of parameters, the seed solutions $\boldsymbol{\xi}_{\epsilon_1}$ and $\overline{\boldsymbol{\xi}}_{-\epsilon_1}$, which form the matrix $U$ in (\ref{U1}), satisfy $\overline{\boldsymbol{\xi}}_{-\epsilon_1}\sim\sigma_2\,\boldsymbol{\xi}_{\epsilon_1}$. This is in line with the construction presented in (\ref{Samsonov})-(\ref{Samsonov2}).

The operator $\mathbf{h_2}$ in (\ref{h2ad-hoc}) reads as
\begin{equation}
\mathbf{h_2}=-i\sigma_1\partial_x+\widetilde{v}_1\,\sigma_0+2(\lambda(x)-\widetilde{v}_1)\mathbb{S}_2,\quad 
\label{h2ad-hoc2}
\end{equation} 
where $\lambda(x)$ is a real function. 
In dependence on its explicit form, the eigenstates of $\mathbf{h_2}$ might not be calculable analytically. Nevertheless the eigenstates of $\mathbf{\widetilde{h}_{1}}$ are known.

Now, we can use (\ref{widetildeHsoc}) to get the Hamiltonian $\widetilde{H}_{soc}$ of spin-orbit interaction.
The electrostatic field $\widetilde{V}$, intrinsic spin-orbit interaction $\widetilde{\Delta}$ and the Rashba term $\widetilde{\lambda}$ read as
\begin{equation}
\widetilde{V}(x)=\frac{1}{2}\lambda(x),\quad \widetilde{\lambda}(x)=\lambda(x),\quad \widetilde{\Delta}(x)=\widetilde{v}_1(x)-\frac{\lambda(x)}{2}.
\end{equation}
The Hamiltonian $\widetilde{H}_{soc}$ is partially-exactly solvable as for each energy $E$, we can find an eigenstate $\widetilde{\boldsymbol{\Phi}}_E=(1,0)^T\otimes\widetilde{\boldsymbol{\xi}}_E$, where
\begin{equation}
(\widetilde{H}_{soc}-E)\widetilde{\boldsymbol{\Phi}}_E=0,\quad (\mathbf{\widetilde{h}_1}-E)\widetilde{\boldsymbol{\xi}}_E=0.
\end{equation}
Notice that these states are independent on the actual choice of $\lambda(x)$.
In particular, there are two localized states,
\begin{equation}\widetilde{\boldsymbol{\Phi}}_{\epsilon_1}=(1,0)^T\otimes\widetilde{\boldsymbol{\xi}}_{\epsilon_1},\quad \widetilde{\boldsymbol{\Phi}}_{-\epsilon_1}=(1,0)^T\otimes\widetilde{\boldsymbol{\xi}}_{-\epsilon_1},\end{equation}
where  $\boldsymbol{\widetilde{\xi}}_{\pm\epsilon_1} $ can be obtained from (\ref{missing-2x2}) together with (\ref{soc-parameters}). 

Let us consider the particular case $\lambda(x)=\widetilde{v}_1(x)$ in $\mathbf{h_2}$. This leads to the eigenvalue problem 
\begin{equation}
\mathbf{h_2}\boldsymbol{\chi}_{E}=(-i\sigma_1\partial_x+\widetilde{v}_1(x)\sigma_0)
\boldsymbol{\chi}_{E}
= E\,
\boldsymbol{\chi}_{E}\label{h2spec}
\end{equation}
the solutions of which can be determined with ease and are given by
\begin{equation}
\boldsymbol{\chi}_{E}=e^{-i\sigma_1 \int_x \widetilde{v}_1(s)ds}\left(c_1\,e^{iE\, x}(1,1)^T+c_2\,e^{-iE\, x}(1,-1)^T\right),
\quad c_{0}, c_{1}\in\mathbb{C}. 
\label{chispec}
\end{equation}
The Hamiltonian $\mathbf{h_2}$ is also reflectionless; indeed, it is known fact that electrostatic barriers are penetrated by Dirac fermions of any energy without back-scattering. This phenomenon is called Klein tunneling. The formula (\ref{chispec}) suggests a simple explanation of this phenomenon  \cite{JakubskyNietoPlyushchay}: the Hamiltonian (\ref{h2spec}) is unitarily equivalent to free particle,
\begin{equation}
\mathbf{h_2}=e^{-i\sigma_1\int_x\widetilde{v}_1(s)ds}(-i\sigma_1\partial_x)e^{i\sigma_1\int_x\widetilde{v}_1(s)ds}.
\end{equation}
Therefore, the corresponding Hamiltonian of spin-orbit coupling $\widetilde{H}_{soc}$, whose components of the potential are
\begin{equation}
\widetilde{V}(x)=\frac{\widetilde{v}_1(x)}{2},\quad \widetilde{\lambda}(x)=\widetilde{v}_1(x),\quad \widetilde{\Delta}(x)=\frac{\widetilde{v}_1(x)}{2}
\end{equation}
is exactly solvable and reflectionless.

\section{Non-Hermiticity and non-reducible Darboux transformation}
\label{sec:non-red}

In this section, we briefly discuss some of the issues that appear when dealing with generic $4\times4$ Darboux transformation, non-hermiticity of the new Hamiltonian in particular. For illustration, we fix the Hamiltonian $H$ in the following manner, 
\begin{equation}
H=\mathbb{S}_1\otimes\mathbf{h_1}+\mathbb{S}_2\otimes \mathbf{h_2}.
\end{equation}
Here, $\mathbf{h_1}$ and $\mathbf{h_2}$ are generic hermitian operators as defined in (\ref{h_a}).

Let us consider the most general matrix $U$ in the following, non-diagonal, block form:
\begin{equation}
U=
\begin{pmatrix}
{U}_{1} & {U}_{3} \\
{U}_{4} & {U}_{2}
\end{pmatrix}
\end{equation}
with ${U}_{j}$, for $j=1,2,3,4$, such that they satisfy
\begin{equation}
\mathbf{h_1}U_1=U_1\Lambda_1,\quad \mathbf{h_1}U_3=U_3\Lambda_2,\quad \mathbf{h_2}U_4=U_4\Lambda_1,\quad \mathbf{h_2}U_2=U_2\Lambda_2,
\end{equation}
where $\Lambda_1=\operatorname{diag}(\epsilon_{1},\epsilon_{2})$ and $\Lambda_1=\operatorname{diag}(\epsilon_{3},\epsilon_{4})$ are constant $2\times2$ diagonal matrices. 
There holds
\begin{equation}
HU=U\begin{pmatrix}
{\Lambda}_{1} & \mathbf{0} \\
\mathbf{0} & {\Lambda}_{2}
\end{pmatrix} 
\end{equation}
with $\mathbf{0}$ the null $2\times 2$ matrix, so that $U$ leads to the Hamiltonian $\widetilde{H}=\gamma\partial_{x}+\widetilde{{V}}$, with $\widetilde{{V}}={V}+i[\gamma,U_{x}U^{-1}]$, $\gamma=-i\sigma_{0}\otimes\sigma_{1}$, and ${V}^{\dagger}={V}$. To this end, we conveniently rewrite the inverse of $U$ in a non-diagonal block form, $U^{-1}=\begin{pmatrix} \overline{U}_{1} & \overline{U}_{3} \\ \overline{U}_{4} & \overline{U}_{2} \end{pmatrix}$, where $UU^{-1}=\mathbb{I}$ and
\begin{equation}
\begin{aligned}
\overline{U}_{1}=(U_{1}-U_{3}U_{2}^{-1}U_{4})^{-1}, \quad \overline{U}_{3}=(U_{4}-U_{2}U_{3}^{-1}U_{1})^{-1}, \\ \overline{U}_{4}=(U_{3}-U_{1}U_{4}^{-1}U_{2})^{-1}, \quad \overline{U}_{2}=(U_{2}-U_{4}U_{1}^{-1}U_{3})^{-1}.
\end{aligned}
\end{equation}
In this form, we may determine the most general expression for the matrix potential as
\begin{equation}
\widetilde{{V}}={V}+i
\begin{pmatrix}
[\sigma_{1},(\partial_{x}U_{1})\overline{U}_{1}+(\partial_{x}U_{3})\overline{U}_{4}] & [\sigma_{1},(\partial_{x}U_{1})\overline{U}_{3}+(\partial_{x}U_{3})\overline{U}_{2}] \\
[\sigma_{1},(\partial_{x}U_{4})\overline{U}_{1}+(\partial_{x}U_{2})\overline{U}_{4}] & [\sigma_{1},(\partial_{x}U_{4})\overline{U}_{3}+(\partial_{x}U_{2})\overline{U}_{2}]
\end{pmatrix},
\end{equation}
provided that all matrices $U_{j}$ are invertible.
The latter formula suggests that ${\widetilde{V}}$ is hermitian as long as the diagonal blocks are hermitian, whereas the anti-diagonal blocks must be one the adjoint of the other. Those conditions are quite restrictive and cannot be guaranteed in general.
	
As a particular example to illustrate non-Hermitian constructions, let us consider the $4\times 4$ free particle Hamiltonian $H=\mathbb{S}_{1}\otimes\widetilde{\boldsymbol{h}}_{1}+\mathbb{S}_{2}\otimes\widetilde{\boldsymbol{h}}_{2}$, with $\boldsymbol{\widetilde{h}}_{j}$ given in~\eqref{widetildeh1h2}, together with $U_{4}=\mathbf{0}$. The latter choice is always feasible as the null matrix corresponds to the trivial solution to the eigenvalue equation associated with $U_{4}$. The remaining matrices $U_{1}$, $U_{2}$, and $U_{3}$ are defined in terms of the general one introduced in~\eqref{U1} through the reparametrizations
\begin{equation}
U_{1}=U\vert_{\substack{v\rightarrow v_{1}, a\rightarrow a_{1}, w\rightarrow w_{1}\\ \epsilon_{1}\rightarrow \epsilon_{1}, \epsilon_{2}\rightarrow\epsilon_{2} \\ \delta_{1}\rightarrow\delta_{1}, \delta_{2}\rightarrow\delta_{2}}} \, , \quad 
U_{2}=U\vert_{\substack{v\rightarrow v_{2}, a\rightarrow a_{2}, w\rightarrow w_{2}\\ \epsilon_{1}\rightarrow \epsilon_{3}, \epsilon_{2}\rightarrow\epsilon_{4} \\ \delta_{1}\rightarrow\delta_{3}, \delta_{2}\rightarrow\delta_{4}}} \, , \quad 
U_{3}=U\vert_{\substack{v\rightarrow v_{2}, a\rightarrow a_{2}, w\rightarrow w_{2}\\ \epsilon_{1}\rightarrow \epsilon_{3}, \epsilon_{2}\rightarrow\epsilon_{4} \\ \delta_{1}\rightarrow\overline{\delta}_{3}, \delta_{2}\rightarrow\overline{\delta}_{4}}} \, .
\label{U1U2U3}
\end{equation}
In this case, we get $\operatorname{det}(U) = \operatorname{det}(U_{1}) \operatorname{det}(U_{2})$, where the individual $2\times 2$ determinants can be each proved to be different from zero. See discussion in Sec.~\ref{sec:2x2-reflectionless}. In this way, $U$ is invertible and takes the form
\begin{equation}
U^{-1}=
\begin{pmatrix}
U_{1}^{-1} & -U{1}^{-1}U_{3}U_{2}^{-1} \\
\mathbf{0} & U_{2}^{-1}
\end{pmatrix}
\, , \quad
(U^{-1})^{\dagger}=
\begin{pmatrix}
(U_{1}^{-1})^{\dagger} & \mathbf{0} \\
-(U_{2}^{-1})^{\dagger}U_{3}^{\dagger}(U_{1}^{-1} )^{\dagger} & (U_{2}^{-1})^{\dagger}
\end{pmatrix}
\, .
\label{inverseU}
\end{equation}
whereas the Hamiltonian becomes
\begin{equation}
\widetilde{H}=H+
i
\begin{pmatrix}
[\sigma_{1},(\partial_{x}U_{1})U_{1}^{-1}] & [\sigma_{1},(\partial_{x}U_{3})U_{2}^{-1}-(\partial_{x}U_{1})U_{1}^{-1}U_{3}U_{2}^{-1}] \\ 
\mathbf{0} & [\sigma_{1},(\partial_{x}U_{2})U_{2}^{-1}] 
\end{pmatrix} \, .
\label{U-nonsep}
\end{equation}
In the latter, the diagonal blocks correspond to the potential matrix generated by $2\times 2$ Darboux transformation, see (\ref{VV1}), which are hermitian. Nevertheless, the non-vanishing upper-antidiagonal block breaks hermiticity of the Hamiltonian $\widetilde{H}$. We can  impose the condition
\begin{equation*}
(\partial_{x}U_{3})U_{2}^{-1}-(\partial_{x}U_{1})U_{1}^{-1}U_{3}U_{2}^{-1}=f_{0}(x)\sigma_{0}+f_{1}(x)\sigma_{1} \, ,
\end{equation*}
with $f_{0}(x)$ and $f_{1}(x)$ some functions, possibly null, to be determined. Nevertheless, fixing hermiticity of $\widetilde{H}$ this way would make the  Hamiltonian reducible. Therefore, by departing from a separable Hamiltonian, it seems that a tight relationship exists between the hermiticity and the separability of the  Hamiltonian obtained via the Darboux transformation. Thus, the concept of reducible Darboux transformations applied on reducible Hamiltonians gives us an additional insight on how to construct appropriate Darboux transformations such that the new model is free of singularities and hermitian, a task not so clear from the non-separable setup. 

\begin{figure}
	\centering
	\includegraphics[width=0.35\textwidth]{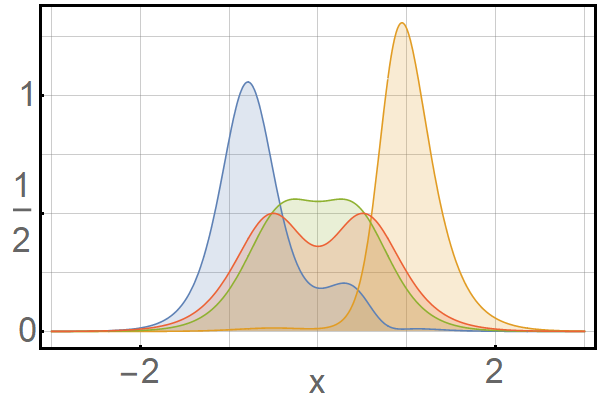}
	\caption{Probability densities $\overline{\mathcal{P}}_{\epsilon_{j}}(x)$ associated with the missing eigensolutions $\overline{\boldsymbol{\Phi}}_{\epsilon_{j}}$ for $\epsilon_{1}$ (blue), $\epsilon_{2}$ (orange), $\epsilon_{3}$ (green), and $\epsilon_{4}$ (red). The parameters, introduced in~\eqref{U1U2U3}, have been fixed as $v_{1}=3$, $v_{2}=2.5$, $w_{1}=-2$, $w_{2}=-1.5$, $a_{1}=i$, $a_{2}=1$, $\epsilon_{1}=1.25$, $\epsilon_{2}=0.25$, $\epsilon_{3}=0.75$, $\epsilon_{4}=-0.5$, $\delta_{1}=1$, $\delta_{2}=-1$, and $\delta_{3}=\overline{\delta}_{3}=\delta_{4}=\overline{\delta}_{4}=0$}
	\label{fig:PD-nonsep}
\end{figure}

Let us briefly discuss some aspects of Darboux transformation when $\widetilde{H}$ is non-Hermitian. To begin with, it is worth noticing that, although the intertwining relation~\eqref{intert1} still holds true when $\widetilde{H}$ is non-Hermitian, its adjoint ceases to related $H$ and $\widetilde{H}$ as we get instead $\mathcal{L}^{\dagger}\widetilde{H}^{\dagger}=H\mathcal{L}^{\dagger}$. That is, $\mathcal{L}^{\dagger}$ maps eigensolutions of $\widetilde{H}^{\dagger}$ into eigensolutions of $H$. Since $(U^{-1})^{\dagger}$ is annihilated by $\mathcal{L}^{\dagger}$, it provides us with the missing eigenstates of $\widetilde{H}^{\dagger}$. Explicitly, we have
\begin{equation}
\widetilde{H}^{\dagger}\overline{\boldsymbol{\Phi}}_{\epsilon_{k}}=\epsilon_{k}\overline{\boldsymbol{\Phi}}_{\epsilon_{k}}, \quad (U^{-1})^{\dagger}\propto(\overline{\boldsymbol{\Phi}}_{\epsilon_{1}},\ldots,\overline{\boldsymbol{\Phi}}_{\epsilon_{4}}), \quad k\in\{1,2,3,4\}.
\end{equation} 
Comparing with the explicit form of $U$ in~\eqref{inverseU}, we can see that the missing eigenstates $\overline{\boldsymbol{\Phi}}_{\epsilon_{3}}$ and $\overline{\boldsymbol{\Phi}}_{\epsilon_{4}}$ coincide with $\overline{\boldsymbol{\Phi}}_{\epsilon_{3}}$ and $\overline{\boldsymbol{\Phi}}_{\epsilon_{4}}$ of the separable case in Sec.~\ref{Hsocreflectionless}. The difference emerges in the eigensolutions $\overline{\boldsymbol{\Phi}}_{\epsilon_{1}}$ and $\overline{\boldsymbol{\Phi}}_{\epsilon_{2}}$ that get modified by presence of $U_{3}$ in $U$. Such a behavior is depicted in the probability densities presented in Fig.~\ref{fig:PD-nonsep}, where it is clear that only $\overline{\mathcal{P}}_{\epsilon_{3}}$ (green) and $\overline{\mathcal{P}}_{\epsilon_{4}}$ (red) are the same as their separable counterparts presented in Fig.~\ref{dist-fig}.

It is worth to mention that, despite the lack of Hermiticity, the eigenvalues added to the new Hamiltonian $\widetilde{H}^{\dagger}$ are all-real, and their corresponding eigenfunctions have finite-norm. Here, we shall not discuss the properties mathematical apparatus behind non-Hermitian structures such as pseudo-Hermiticity, metric operators, bi-orthogonality, or PT-symmetry, as those deserve attention by their own. Rather, by means of this particular example, we want to point out the apparent connection between Hermiticity and separability.

\section{Concluding remarks\label{section7}}

In the article, we discussed possible problems and prospects related with the use of Darboux transformation in the context of physical systems described by $4\times4$ Dirac Hamiltonians. We reviewed the general framework of Darboux transformation in section \ref{sec:Dirac}, showing explicitly the problems that appear when we require it to produce energy operators of specific form. As follows from (\ref{hermitianregular}) and (\ref{identification}) , the difficulty to get the new operator in the required form increases rapidly with dimension of the involved matrices. In this sense, the $2\times 2$ Darboux transformation is the easiest to deal with. In section \ref{section3}, we focused on a specific class of $4\times4$ Dirac operator that are reducible, i.e. they can be brought into block-diagonal form by a unitary transformation (\ref{H-blocks}). We showed that both the Hamiltonians of distortion scattering and spin-orbit interaction belong to this class, see (\ref{Hdisred}) and (\ref{Hsocred}). This observation paves the way to use $2\times2$ Darboux transformation in construction of these $4\times4$ Dirac Hamiltonians such that the form of the potential term is granted by construction. In section \ref{sec:2x2-reflectionless}, we discuss in detail $2\times2$ Darboux of free-particle system. We derive a class of systems that posses bound states and are reflectionless. We used these results in section  \ref{section5}, where reflectionless Hamiltonians with spin-orbit coupling and distortion scattering are constructed. The section \ref{sec:non-red} was devoted to discussion of non-reducible Darboux transformations.  

It is worth mentioning that both the Hamiltonian of distortion scattering and of spin-orbit interaction in (\ref{Hsoc}) are just two members of the family of reducible Dirac operators. The concept of reducibility as reflected in (\ref{H-blocks}) has much broader applicability. We also found that the reducibility allows us to define \textit{partial} Darboux transformations (\ref{red-Lsoc}) that lie out of the usual definition (\ref{L}). We believe that these concepts and their application in description of physical systems are worth of further investigation, which, however, goes beyond the scope of the current article.



\section*{Acknowledgment}
M.C.-C. thanks Department of Physics of the Nuclear Physics Institute of CAS for hospitality. M.C.-C. acknowledges the support of CONACYT, project FORDECYT-PRONACES/61533/2020. M.C.-C. also acknowledges the Conacyt fellowship 301117. V. J. was supported by GACR grant no 19-07117S. K.Z. acknowledges the support from the project “Physicists on the move II” (KINE\'O II), Czech Republic, Grant No. CZ.02.2.69/0.0/0.0/18 053/0017163.

\appendix
\setcounter{section}{0}
\section*{Appendix }
\label{appA}
\renewcommand{\thesection}{A-\arabic{section}}
\renewcommand{\theequation}{A-\arabic{equation}}
\setcounter{equation}{0}  

We shall find asymptotic form of the matrix $U_xU^{-1}$, where $U$ is given in (\ref{U1}). 
 We have
\begin{equation}
(\partial_xU)U^{-1}=\frac{1}{\mathcal{D}(x)}\left(\begin{matrix}
f_{11} & f_{12} \\
f_{21} & f_{22}
\end{matrix}\right),    
\end{equation}
where the matrix elements and the denominator are as follows 
\begin{equation}
f_{11}=ic_1+c_2\tanh(z_1)-i\,a^*\,\kappa_{\epsilon_2}(\text{Im}\,a+\kappa_{\epsilon_1}\tanh(z_1))\tanh({z}_2),
\end{equation}
\begin{equation}
f_{22}=i\widetilde{c}_1+\widetilde{c}_2\tanh({z}_2)+i\,a\,\kappa_{\epsilon_1}(\text{Im}\,a-\kappa_{\epsilon_2}\tanh({z}_2))\tanh(z_1),
\end{equation}
\begin{equation}
f_{12}=i (w-\epsilon_1) (\kappa_{\epsilon_2}^2+\text{Im}\,a\, \kappa_{\epsilon_1} \tanh (z_1)-\kappa_{\epsilon_2} (\text{Im}\,a+\kappa_{\epsilon_1} \tanh (z_1))\tanh ({z}_2)),
\end{equation}
\begin{equation}
f_{21}=i (v-\epsilon_2) (\kappa_{\epsilon_1}^2+ \text{Im}\,a\,\kappa_{\epsilon_1}\tanh(z_1)-\kappa_{\epsilon_2}(\text{Im}\,a+\kappa_{\epsilon_1} \tanh (z_1)) \tanh ({z}_2)).
\end{equation}
and $\mathcal{D}(x)$ is given in (\ref{detU2}). For the sake of completeness we put the formula here again
\begin{equation}
\mathcal{D}(x)=(v-\epsilon_{2})(w-\epsilon_{1})+\left( \kappa_{\epsilon_{1}}\tanh(z_{1})+\operatorname{Im}a\right)\left( \kappa_{\epsilon_{2}}\tanh({z}_{2})-\operatorname{Im}\,a\right) .
\end{equation}
The $c_1$ and $c_2$ constants are given as
\begin{equation}
\begin{split}
& c_1=\text{Im}\,a^2\text{Re}\,a-i\text{Im}\,a\kappa_{\epsilon_2}^2-\text{Re}\,a(w-\epsilon_1)(v-\epsilon_2),\\
& c_2=i\text{Im}\,a\text{Re}\,a+\kappa_{\epsilon_2}^2+(w-\epsilon_1)(v-\epsilon_2),
\end{split}
\end{equation}
\begin{equation}
\widetilde{c}_1=c_1^*|_{\kappa_{\epsilon_2}\rightarrow\kappa_{\epsilon_1}},\quad \widetilde{c}_2=c_2^*|_{\kappa_{\epsilon_2}\rightarrow\kappa_{\epsilon_1}}.
\end{equation}
Now, we calculate the limit $x\rightarrow\pm\infty$ of the matrix $(\partial_xU)U^{-1}$, 
\begin{equation}
\begin{split}
&w_{\pm}=\lim_{x\rightarrow\pm\infty} (\partial_xU)U^{-1}=\\
&\frac{1}{\mathcal{D}^\pm}\left(\begin{matrix}
ic_1\pm c_2\mp i\,a^*\kappa_{\epsilon_2}(\text{Im}\,a\pm\kappa_{\epsilon_1}) & i (w-\epsilon_1) (\kappa_{\epsilon_2}^2\pm\text{Im}\,a \kappa_{\epsilon_1} \mp\kappa_{\epsilon_2} (\text{Im}\,a\pm\kappa_{\epsilon_1})) \\
i (v-\epsilon_2) (\kappa_{\epsilon_1}^2\pm \text{Im}\,a\kappa_{\epsilon_1}\mp\kappa_{\epsilon_2}(\text{Im}\,a\pm\kappa_{\epsilon_1})) & i\widetilde{c}_1\pm\widetilde{c}_2\pm i\,a\,\kappa_{\epsilon_1}(\text{Im}\,a\mp\kappa_{\epsilon_2})
\end{matrix}\right),    
\end{split}
\end{equation}
where 
\begin{equation*}
\mathcal{D}^\pm=-\text{Im}{a}^2\pm\kappa_{\epsilon_2}(\text{Im}\,a\pm\kappa_{\epsilon_1} )\mp\text{Im}\,a \kappa_{\epsilon_1}+(v-\epsilon_2) (w-\epsilon_1).
\end{equation*}


\end{document}